\documentclass[twocolumn]{aastex631}
\usepackage{amsmath,amssymb}
\usepackage{natbib}
\usepackage{graphicx}
\usepackage[figuresright]{rotating}
\usepackage{threeparttable}
\usepackage{multirow}
\usepackage{booktabs}
\usepackage{hyperref}

\newcommand{\OIII}{[O\,{\sc iii}]}
\newcommand{\OIIIlam}[1]{[O\,{\sc iii}]$\lambda${#1}}
\newcommand{\OII}{[O\,{\sc ii}]}
\newcommand{\NII}{[N\,{\sc ii}]}

\newcommand{\CII}{[C\,{\sc ii}]}
\newcommand{\Ha}{H$\alpha$}
\newcommand{\Hb}{H$\beta$}
\newcommand{\Hg}{H$\gamma$}

\newcommand{\NAOJ}{National Astronomical Observatory of Japan, 2-21-1 Osawa, Mitaka, Tokyo 181-8588, Japan}
\newcommand{\SOKENDAI}{Department of Astronomical Science, The Graduate University for Advanced Studies, SOKENDAI, 2-21-1 Osawa, Mitaka, Tokyo 181-8588, Japan}
\newcommand{\UA}{Department of Astronomy and Steward Observatory, University of Arizona, Tucson, AZ 85721, USA}
\newcommand{\ICRR}{Institute for Cosmic Ray Research, The University of Tokyo, 5-1-5 Kashiwanoha, Kashiwa, Chiba 277-8582, Japan}
\newcommand{\UT}{Department of Physics, Graduate School of Science, The University of Tokyo, 7-3-1 Hongo, Bunkyo, Tokyo 113-0033, Japan}
\newcommand{\TSUKUBA}{Center for Computational Sciences, University of Tsukuba, Ten-nodai, 1-1-1 Tsukuba, Ibaraki 305-8577, Japan}

\shortauthors{Nishigaki et al.}
\graphicspath{{./}{figures/}}

\begin{document}

\title{
DREAMS. II.\\
Galaxy Demographics from Direct $T_{\rm e}$-Based Metallicities at $z\sim 2–10$:\\
Tracing the Evolution of the Mass-Metallicity and Fundamental Relations
}

\author[0000-0003-4321-0975]{Moka Nishigaki}
\affiliation{\NAOJ}\affiliation{\SOKENDAI}
\author[0000-0003-2965-5070]{Kimihiko Nakajima}
\affiliation{Institute of Liberal Arts and Science, Kanazawa University, Kakuma-machi, Kanazawa, Ishikawa, 920-1192, Japan}\affiliation{\NAOJ}
\author[0000-0002-1049-6658]{Masami Ouchi}
\affiliation{\NAOJ}\affiliation{\ICRR}
\affiliation{Astronomical Science Program, Graduate Institute for Advanced Studies, SOKENDAI, 2-21-1 Osawa, Mitaka, Tokyo 181-8588, Japan}
\affiliation{Kavli Institute for the Physics and Mathematics of the Universe (Kavli IPMU, WPI), The University of Tokyo, 5-1-5 Kashiwanoha, Kashiwa, Chiba, 277-8583, Japan}
\author[0000-0002-2517-6446]{Peter Behroozi}
\affiliation{\UA}
\author[0009-0000-1999-5472]{Minami Nakane}
\affiliation{\ICRR}\affiliation{\UT}
\author[0009-0005-2897-002X]{Yui Takeda}
\affiliation{\NAOJ}\affiliation{\SOKENDAI}
\author[0009-0008-0167-5129]{Hiroya Umeda}
\affiliation{\ICRR}\affiliation{\UT}
\author[0000-0002-1319-3433]{Hidenobu Yajima}
\affiliation{\TSUKUBA}
\author[0009-0006-6763-4245]{Hiroto Yanagisawa}
\affiliation{\ICRR}\affiliation{\UT}
\begin{abstract} 
We present the statistics of line ratios and direct $T_{\rm e}$-based metallicities
from JWST medium-resolution spectra of 292 galaxies at $z=2$–10, combining DREAMS observations with those of JADES and CEERS.
To remove systematics caused by stellar mass ($M_*$) and star formation rate (SFR), we construct stacked spectra binned by redshift within fixed $M_*$ and SFR ranges, as well as across the full ranges.  We find that the \OII$\lambda$3727/\Hb\ ratio drops by a factor of five from $z\sim 3$ to 8 at fixed $M_*$ and SFR, in contrast to the nearly constant \OIII$\lambda$5007/\Hb\ ratio. 
We derive metallicities via the direct $T_{\rm e}$ method using the \OIII$\lambda$4363 line, and identify that high-$z$ galaxies lie on the low-metallicity end of the anti-correlation between ionization parameter and metallicity at $z\sim0$.
Photoionization modeling demonstrates that the redshift evolution, where metallicity decreases and ionization parameter increases, self-consistently explains the observed line ratios. We then examine the mass–metallicity ($M_*$–$Z$) and fundamental ($M_*$–$Z$–SFR) relations. Including additional galaxies at $z\sim 10-12$, we find that the $M_*$–$Z$ relation monotonically decreases from $z\sim 3$ to 10 at fixed $M_*$, while the $M_*$–$Z$–SFR relation shows
a significant decline at $z\gtrsim8$.
Based on the \textsc{ChemicalUniverseMachine} model, this evolutionary trend can be explained by enhanced gas inflow (outflow) by a factor of $\sim 5$ ($\sim 1.7$) at $z\gtrsim8$.
\end{abstract}
\section{Introduction} \label{sec:intro}
The chemical composition of galaxies encodes essential information about their formation and evolutionary histories. In particular, the gas-phase metallicity, the abundance of heavy elements in the interstellar medium (ISM), serves as a fossil record of past star formation and the interplay between galaxies and their environments. As stars form and evolve, they enrich the surrounding gas with metals through stellar winds and supernova explosions. At the same time, galaxies grow through gas inflows from the intergalactic medium (IGM) and lose metals via outflows driven by stellar or AGN feedback. The metallicity thus reflects a balance between these competing processes and provides key constraints on the so-called baryon cycle.

A major step forward in characterizing this balance came with the discovery of the Fundamental Metallicity Relation \citep[FMR;][]{Mannucci10,Maiolino19}, which links metallicity not only to stellar mass but also to star formation rate (SFR). First established in the local universe, the FMR suggests that, at fixed stellar mass, galaxies with higher SFRs tend to have lower metallicities. This three-dimensional relation has been interpreted as a manifestation of the equilibrium between gas accretion, star formation, and metal-enriched outflows \citep[e.g.,][]{Finlator08,Dave12,Lilly13}. Because the FMR encodes the physical conditions under which galaxies grow, its existence and possible evolution at high redshifts provide critical insights into how star formation and feedback operated in the early universe.
These observational probes provide a crucial test for modern cosmological simulations, which predict a gradual evolution of the FMR driven by factors such as higher gas fractions and more intense feedback at early cosmic times \citep[e.g.,][]{Ma16,Dave17,Torrey19}.

Until recently, probing the FMR at high redshifts had been limited by observational constraints, particularly the difficulty of accessing rest-frame optical emission lines beyond $z \sim 3$ with ground-based telescopes. The James Webb Space Telescope (JWST), with its unprecedented near-infrared sensitivity and spectroscopic capabilities, has revolutionized this field by enabling detailed studies of galaxies out to $z \sim 10$. 
\cite{Nakajima23} used JWST/NIRSpec data from the ERO and CEERS survey to investigate the mass–metallicity and FMR relations in galaxies at $z = 4$–10. By combining strong-line estimates (primarily using the R23 index, defined as 
[\OIII$\lambda\lambda$4959,5007 $+$ \OII$\lambda$3727]/\Hb) 
for the majority of their sample with direct electron temperature ($T_\mathrm{e}$) measurements from detected \OIII$\lambda$4363 in ten galaxies, 
they find consistency with the local FMR at $z \sim 4$--$6$, but suggest a significant offset at $z \gtrsim 8$.
In parallel, \cite{Curti24} analyzed JADES DR1 spectra of galaxies at $z = 3$–10, deriving metallicities using a combination of multiple strong-line calibrations including \OIII$\lambda5007$/\Hb\ (R3), and similarly found deviations from the local FMR emerging at $z \gtrsim 6$. 
These studies reveal tentative signs that the FMR may begin to break down at $z \gtrsim$ 6--8, where galaxies appear to be significantly more metal-poor than expected based on their stellar mass and SFR alone.

However, the difference results from the adopted different definitions of the FMR. \cite{Nakajima23} employed the formulation of \cite{AM13}, while \cite{Curti24} used another calibration based on \cite{Curti20}. Since the predicted metallicity depends sensitively on how the FMR is parameterized (i.e., the functional dependence on $M_*$ and SFR), these differences in definition complicate the direct comparison of high-redshift data to the local relation. This highlights the importance of analyzing the evolution of metallicity in a manner that does not depend on a specific FMR definition.

Moreover, strong-line metallicity indicators, such as the R23 or O32 indices, are calibrated using local galaxy samples and empirical or photoionization-based relations. While these diagnostics are widely used due to their accessibility, they are subject to systematic uncertainties and may not accurately reflect the ISM conditions in early galaxies, which are expected to differ from their low-redshift counterparts in terms of ionization state, gas density, and elemental abundance ratios. A more direct and physically motivated approach is the so-called direct $T_\mathrm{e}$ method, which derives the electron temperature from the weak auroral \OIII$\lambda$4363 line and uses it to compute the oxygen abundance. Although this method provides more reliable metallicity estimates, it is challenging to apply at high redshift because the required emission lines are often too faint to detect in individual spectra.

The ionization parameter also provides key diagnostic of the physical conditions in star-forming gas. The ionization parameter depends on the ISM properties such as the hardness of the ionizing radiation field, the gas density, and the geometry of the ionized regions, and it strongly influences the relative strengths of emission lines used for abundance measurements. Large spectroscopic samples from SDSS at $z \sim 0$ suggest an empirical anti-correlation between ionization parameter and gas-phase metallicity, in which metal-poor galaxies tend to have higher ionization parameters. Similar trends have been suggested at $z \sim 2$–3 from studies of individual galaxies \citep{Nakajima14,Sanders20}. However, the relation has not yet been robustly tested at $z \gtrsim 4$. 
While initial JWST studies suggest that early galaxies indeed have elevated ionization parameters \citep[e.g.,][]{Cameron23,Sanders23,Topping24}, a systematic investigation across a wide redshift range is still lacking.
Measuring the typical ionization parameter and its redshift evolution at these early epochs is essential for understanding the excitation properties of high-redshift galaxies and for applying metallicity diagnostics calibrated at lower redshifts.

While stacking techniques have been successfully applied in previous surveys at $z\sim2-3$ to study average ISM properties \citep[e.g.,][]{Steidel14,Sanders15,Sanders20}, the detection of faint auroral lines needed for the $T_\mathrm{e}$ method remained challenging. The unparalleled sensitivity of JWST now allows us to push this technique to higher redshifts and overcome these prior limitations.
In this work, we overcome these limitations by using a powerful combination of JWST/NIRSpec observations. We begin with the DREAMS survey, which targets gravitationally lensed galaxies, and expand our sample with the large number of galaxies available in the JADES and CEERS programs. By stacking spectra from these surveys in redshift bins from $z\sim2$ to $z\sim10$, we achieve a signal-to-noise ratio sufficient to detect the \OIIIlam{4363} auroral line. 
This allows us to derive robust, direct-method metallicities and assess the average ISM properties at each epoch.
We then examine how these metallicities relate to stellar mass and SFR, and evaluate the degree to which high-redshift galaxies deviate from the local FMR. We further investigate the physical properties by dividing our galaxy sample into subsamples based on $M_*$ and SFR. This binning allows us to study the high-redshift FMR in a definition-independent way and to quantify any deviations as a function of physical galaxy properties. 
Finally, we compare our results with predictions from theoretical models to explore the physical mechanisms responsible for the observed evolution of the FMR, and assess whether its apparent breakdown at early times reflects changes in the baryon cycle, such as more efficient gas accretion or suppressed metal retention.

This paper is organized as follows. In Section 2, we describe the data used in this study. Section 3 outlines the methods, including the stacking procedure applied to the spectra. Section 4 presents the results, covering emission line ratios, metallicity measurements, ionization parameters, the mass-metallicity relation (MZR), and the FMR. In Section 5, we discuss the physical implications of the observed evolution of the FMR. Finally, Section 6 summarizes our findings and conclusions.
\section{Data} \label{sec:data}
In this study, we make use of spectroscopic and imaging datasets obtained from multiple JWST programs, including DREAMS, JADES, ERO, and CEERS. These programs collectively provide medium-resolution NIRSpec spectra that cover the key rest-frame optical emission lines, along with complementary NIRCam photometry. The combination of these datasets allows us to construct a large and homogeneous sample of galaxies across a wide redshift range.
\subsection{DREAMS}
\label{sec:dreams}
We make use of observations from the program GO-4750 (PI: K. Nakajima), titled Deep Reconnaissance of Early Assemblies of Metal-poor Star formation (DREAMS). DREAMS targets the lensing cluster field MACS J0416, enabling access to intrinsically faint galaxies through gravitational magnification. The spectroscopic observations were obtained with JWST/NIRSpec using the MSA and the medium-resolution grating configurations G140M/F070LP and G395M/F290LP, with a resolving power of $R \sim 1000$. 
The DREAMS survey and its data reduction are partially described in \cite{Nakajima25}, with a full description to be presented in a forthcoming paper (Nakajima et al. in preparation).
Among the objects observed in DREAMS, five galaxies (IDs 30001, 30002, 30003, 31001, and 40005; Nakajima et al., in preparation) at redshifts $z \sim 8$ cover the full wavelength range from \OII$\lambda3727$ to \OIII$\lambda5007$, so that all key emission lines used for the metallicity determination (i.e., \OII$\lambda3727$, H$\gamma$, \OIII$\lambda4363$, H$\beta$, and \OIII$\lambda5007$) are available.

Photometric data for these objects are taken from other JWST programs. We adopt the reduced catalogs published by \citet{Ma24}, \citet{Willott24}, and \citet{Diego24}. 
Magnification factors are derived from lensing models of MACS J0416 constructed using the techniques described in \citet{Oguri10,Oguri21,Kawamata18}, with field-specific information taken from \citet{Fu25}.
The magnification factors for the DREAMS objects in our sample are 1.38, 1.42, 1.42, 1.05, and 8.4 for IDs 30001, 30002, 30003, 31001, and 40005, respectively.
\subsection{JADES}
We also use both spectroscopic and photometric data from the third data release of the JWST Advanced Deep Extragalactic Survey \citep[JADES;][]{Eisenstein23a,Bunker24,DEugenio25}. This release provides deep imaging and NIRSpec spectroscopy in the GOODS-S and GOODS-N fields, offering the largest number of galaxies in our sample and covering a broad range of redshifts and stellar masses.
The spectroscopic component includes medium-depth and deep observations taken with the NIRSpec microshutter assembly (MSA), covering a spectral range of 0.6–5.3 $\mu$m. These observations employ both the low-resolution prism ($R \sim 30$–300) and all three medium-resolution grating configurations: G140M/F070LP, G235M/F170LP, and G395M/F290LP, which provide spectral resolutions of $R \sim 1000$.
We use the medium-grating spectra to derive spectral features. All one-dimensional spectra were reduced by the JADES team as part of the official data release \citep{DEugenio25}, and spectroscopic redshifts based on these spectra are also provided.

Photometric data are taken from the official JADES photometric catalog, which was also compiled and released by the survey team \citep{Rieke23}. The catalog includes observations in up to 14 filters, depending on the field coverage, including NIRCam bands of F090W, F115W, F150W, F200W, F277W, F356W, and F444W. These broad-band data are essential for deriving stellar population properties such as stellar mass and star formation rate via spectral energy distribution (SED) fitting.

We select galaxies for analysis by cross-matching the photometric and spectroscopic catalogs and excluding sources flagged for data reduction problems (DR\_flag). We further limit our sample to galaxies with spectroscopic redshifts determined from emission lines, retaining only those classified as flag A (based on high-resolution spectra) or flag B (based on low-resolution spectra). 

We also require that each spectrum covers the full wavelength range from \OII$\lambda3727$ to \OIII$\lambda5007$, ensuring that all of the key emission lines used for the metallicity determination (\OII$\lambda3727$, H$\gamma$, \OIII$\lambda4363$, H$\beta$, and \OIII$\lambda5007$) are available. We confirm that these lines are not affected by the instrumental gap of the grating. 
From the JADES catalog, a total of 822 galaxies satisfy these criteria.

\subsection{ERO and CEERS}
Our analysis further incorporates spectra and catalogs provided by \citet{Nakajima23}, which combine observations from the Early Release Observations \citep[ERO;][Proposal ID 2736]{Finkelstein22} and the Cosmic Evolution Early Release Science Survey \citep[CEERS;][Proposal ID 1345]{Finkelstein22}.

The ERO program targeted the SMACS 0723 lensing cluster field. Spectroscopic observations were obtained with JWST/NIRSpec using the MSA and the medium-resolution grating configurations G235M/F170LP and G395M/F290LP, providing a resolving power of $R \sim 1000$. Among the sources in this field, five galaxies at redshifts $z > 5$ show prominent \OIII\ and \Hb\ emission lines, and all five cover the full set of key emission lines used in this study (\OII$\lambda3727$, H$\gamma$, \OIII$\lambda4363$, H$\beta$, and \OIII$\lambda5007$).  
These galaxies have also been observed with JWST/NIRCam in the F090W, F150W, F200W, F277W, F356W, and F444W filters, for which we adopt the photometric catalog described in \citet{Harikane23a}.

The CEERS program provides NIRSpec MSA spectroscopy in the blank field of the EGS. The survey includes both prism ($R \sim 100$) and medium-resolution ($R \sim 1000$) spectra obtained with the G140M, G235M, and G395M gratings. The CEERS medium-resolution spectra catalog contains 74 galaxies at redshifts $z > 4$ with secure detections of \OIII\ and \Hb\ emission lines. From this catalog, we use 43 galaxies that cover the full set of key emission lines used in this study (from \OII$\lambda3727$ to \OIII$\lambda5007$) and have available NIRCam photometry data. CEERS also offers complementary JWST/NIRCam imaging with the F115W, F150W, F200W, F277W, F356W, F410M, and F444W filters, for which we again use the photometric reductions of \citet{Harikane23a}.
\subsection{Galaxies at $z > 9.5$}
\label{sec:high-z}
At redshifts beyond $z > 9.5$, the \OIIIlam{5007} line is redshifted out of the JWST/NIRSpec spectral coverage, which makes it impossible to estimate electron temperatures and thus determine gas-phase metallicities using direct $T_\mathrm{e}$ method. To explore metallicity evolution further into the early universe, we instead compile measurements from the literature.

Table \ref{tab:indiv_9} summarizes a compilation of galaxies at $z > 9.5$ with metallicity estimates derived using alternative methods. These include the $T_e$ method when \OIIIlam{4363} is detected, as well as strong-line diagnostics such as R23, O32, R3, and Ne3O2, depending on the available line detection in each object. The table also includes stellar mass and star formation rate of each galaxy, both of which are derived from SED fitting.

The sample includes five galaxies spanning $z = 9.5$–12.3, and currently represents the earliest known examples of galaxies with individual metallicity measurements. These objects provide crucial benchmarks for probing chemical enrichment during the first $\sim$500 million years of cosmic history.

In our analysis and discussion of chemical evolution at $z > 9.5$, we use the median values of redshift, stellar mass, and star formation rate derived from this compiled sample to ensure consistency and comparability with the lower-redshift population.
\subsection{Emission-line Flux Measurements}
\label{sec:flux_measurement}
We first measure fluxes for individual spectra obtained with medium-resolution gratings. Emission lines, including \Ha, \Hb, \Hg, and \OIII\,$\lambda\lambda$5007,4959, are fitted with Gaussian profiles, with the noise spectrum used to weight the fits. All wavelength measurements are performed on the vacuum wavelength scale. The uncertainty in each flux measurement is estimated by summing the noise levels of spectral bins in quadrature within a $\pm$FWHM  centered on the Gaussian peak. For some objects, the central wavelengths of the fitted Gaussian peaks deviate from those predicted by the JADES catalog redshifts, with offsets of up to $\sim$20 \AA. Most of these redshifts were determined using PRISM spectra, which have lower spectral resolution than the medium-resolution grating data. For the following analysis, we adopt redshifts measured from the Gaussian fitting of the \OIIIlam{5007} line for all objects.

\subsection{Sample Selection}
\label{sec:sample_selection}
For our analysis, we first select galaxies in which both the \Hb\ and \OIII$\lambda5007$ emission lines are detected with signal-to-noise ratios greater than three. 

\begin{figure}[htb]
    \centering
    \includegraphics[width=\linewidth]
        {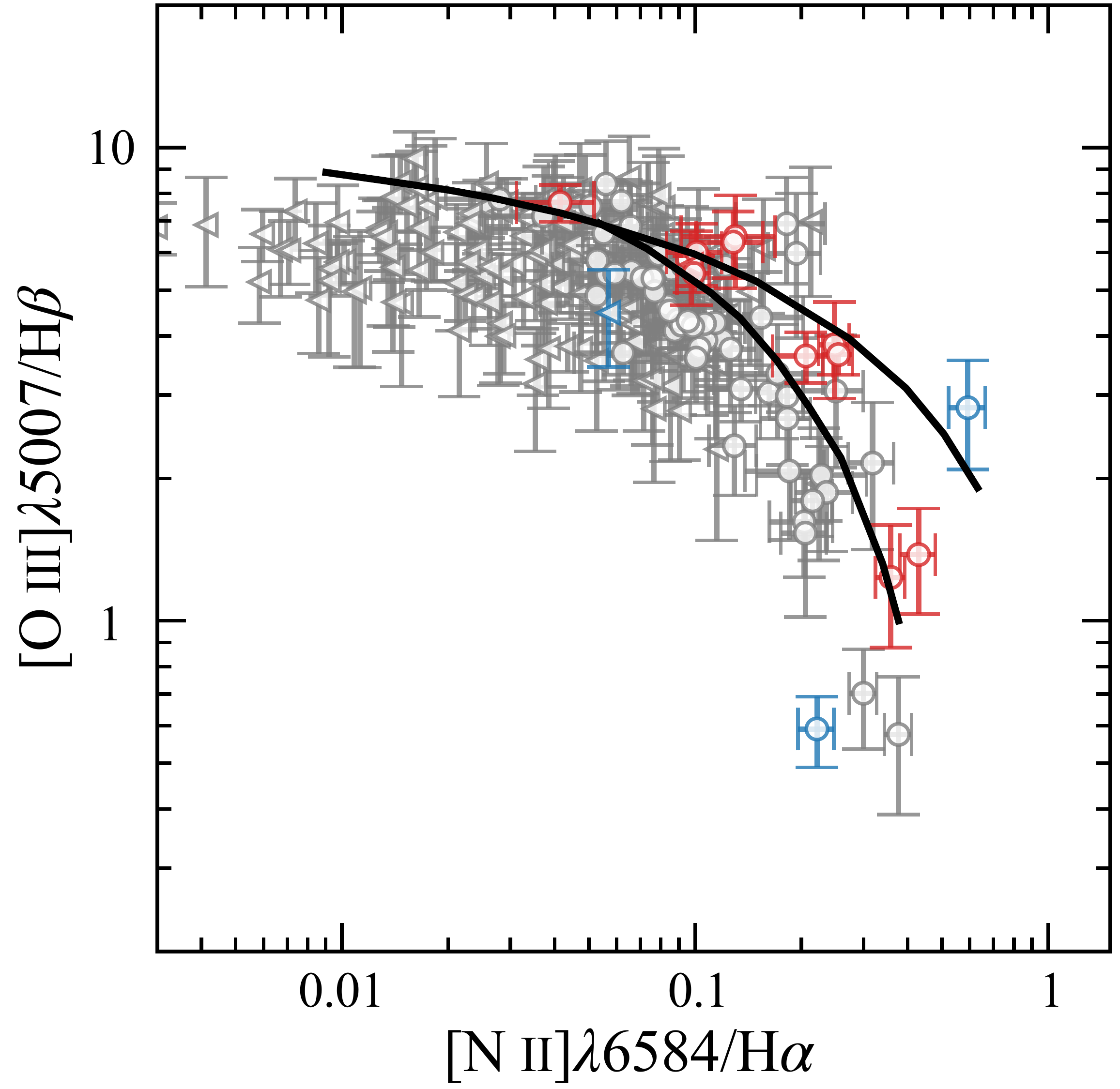}
    \caption{
    The BPT diagram (\OIIIlam{5007}/\Hb vs. \NII$\lambda$6584/\Ha) used for the classification of galaxies and the exclusion of AGN. We plot all galaxies from the parent catalog for which the relevant emission lines are covered and at least one line in each ratio is detected with S/N $>$ 3.
    Circles represent galaxies where all four lines are detected. Triangles indicate 3$\sigma$ upper or lower limits for galaxies where only one line in a given ratio is detected. 
    The solid curves show the AGN/SFG demarcation lines of \citet{Kewley01} (upper) and \citet{Kauffmann03} (lower).
    Galaxies are color-coded according to their classification. 
    Those identified as AGN based on these criteria are shown in red and are excluded from our final sample. Galaxies exhibiting broad H$\alpha$ emission are shown in blue and are excluded from our analysis as broad-line AGN, regardless of their position on the diagram.
    The remaining galaxies, classified as SFGs and used in our analysis, are shown in gray.
    }
    \label{fig:bpt}
\end{figure}

The presence of AGN-driven ionization can bias metallicity calibrations that are optimized for star-forming galaxies and also complicate the determination of stellar masses and star formation rates. To mitigate these effects, we exclude galaxies showing evidence of AGN activity. 
We first remove sources exhibiting broad-line emission in H$\alpha$ or H$\beta$. This includes seven objects in the JADES sample reported by \citet{Maiolino24}, two additional JADES objects we identified (IDs 39435 and 209777), two CEERS objects reported by \citet{Harikane23b} (IDs 00397 and 02782), and one DREAMS object (ID 30001), which will be presented in Takechi et al. (in preparation). 
We also exclude ERO 06355 due to potential AGN activity, which is supported by the detection of [Ne\,{\sc iv}]$\lambda$2423 reported in \citet{Brinchmann23}.
In addition, we apply the classical BPT diagnostic diagram based on the \OIII/H$\beta$ and \NII/H$\alpha$ line ratios as shown in Figure \ref{fig:bpt}. Among galaxies at $z < 6.9$, where H$\alpha$ falls within the NIRSpec wavelength coverage, we remove 15 further objects that satisfy the AGN selection criteria of \citet{Kewley01} and \citet{Kauffmann03} as show in Figure \ref{fig:bpt}. 

In summary, our final sample consists of 292 galaxies, which are used in the following analysis.
The DREAMS survey, though comprising modest sample size of four galaxies, provides a critical contribution to our investigation of the high-redshift frontier. By combining our data, we increase the total number of galaxies at $z>7.8$ with NIRSpec spectroscopy to 11; the physical properties of these individual sources are listed in Table \ref{tab:indiv_9}. For comparison, previous compilations from ERO, GLASS, and CEERS by \citet{Nakajima23} and from JADES DR1 by \citet{Curti24} contained seven and three galaxies in this redshift range, respectively, with uncertainties that made it difficult to clearly discern evolutionary trends. The DREAMS data therefore play a key role by providing robust constraints in a regime where previous studies were limited by large uncertainties.

\section{Method} \label{sec:method}
The aim of this study is to determine the average spectral properties of galaxies at different redshifts and to characterize their typical physical conditions, such as emission line ratios and electron temperature ($T_\mathrm{e}$)-based metallicities. Since \OIIIlam{4363} auroral lines required for direct $T_\mathrm{e}$ metallicity measurements are faint, spectral stacking is essential to enhance the signal-to-noise ratio and enable reliable detection. To minimize selection biases and ensure a meaningful comparison across cosmic time, we also construct a subsample divided within fixed ranges of stellar mass and star formation rate (hereafter, the fixed $M_*$–SFR subsample).
In this section, we describe the procedures used to estimate stellar masses and star formation rates for individual galaxies. We then detail our stacking methodology and the treatment of uncertainties in the derived physical quantities. Finally, we introduce the photoionization models employed to interpret the observed line ratios in terms of underlying physical parameters such as ionization parameters.
\subsection{Stellar Masses and SFRs}
\label{sec:sed_fitting}
\begin{figure*}[htb]
    \centering
    \includegraphics[width=\linewidth]
        {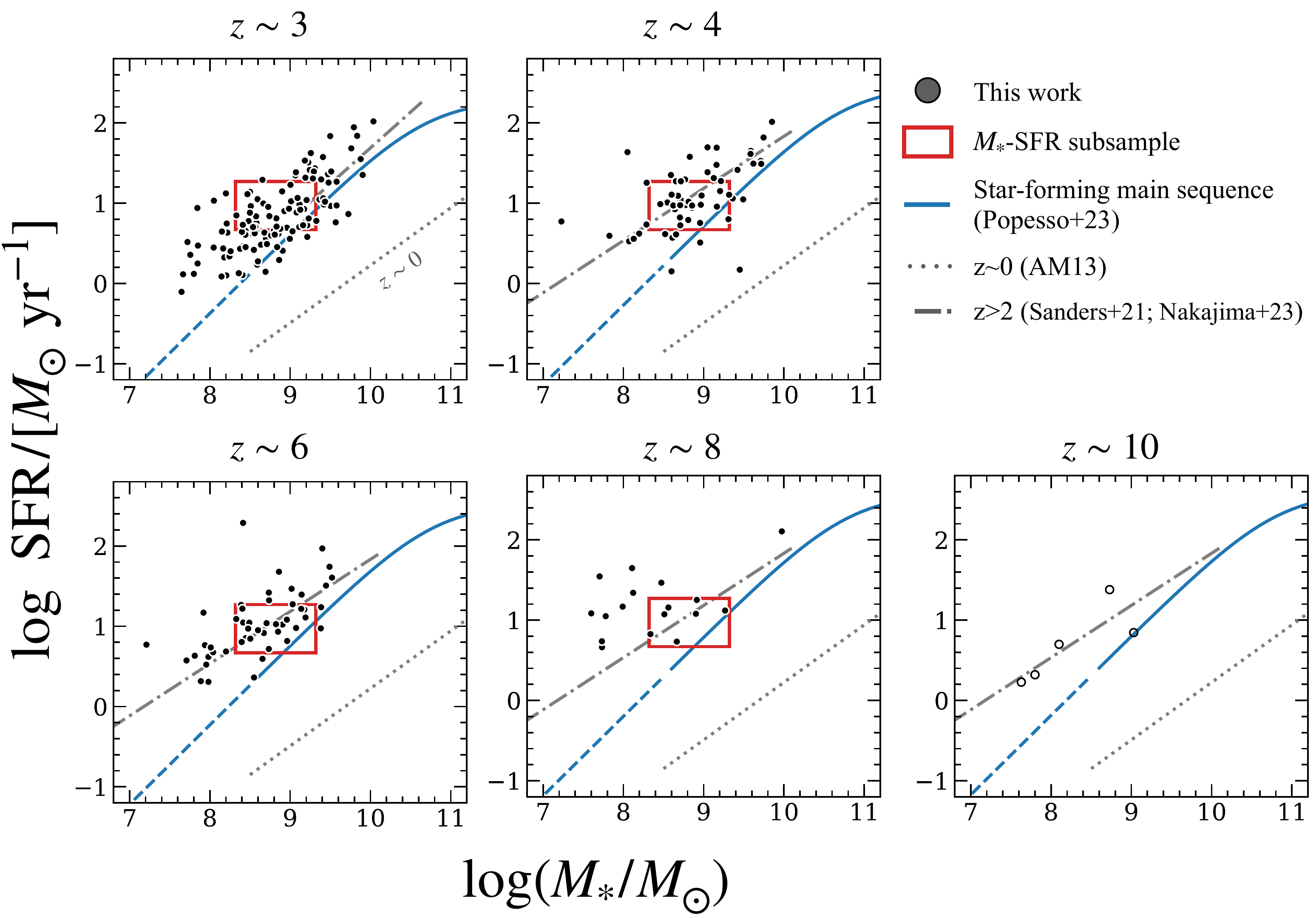}
    \caption{
    Star formation rate versus stellar mass for the galaxy sample used in this study, shown in five redshift bins: $z = 2–4,\, 4–5.5,\, 5.5–7,\, 7–9.5,$ and $z=9.5-12$.
    Filled gray dots represent our sample galaxies, with stellar masses derived from Prospector SED fitting, and SFRs calculated from dust-corrected \Hb\ luminosities.
    In the highest-redshift panel, open gray dots indicate literature data points. Note that the methods used to estimate stellar mass and SFR differ across these studies.
    The blue solid line in each panel shows the predicted star formation main sequence from the simulation by \cite{Popesso23} at the corresponding redshift.
    For comparison, we show samples used in previous FMR studies:
    the gray dotted line shows the local ($z \sim 0$) relation from \cite{AM13}, while the gray dashed and dot-dashed lines correspond to $z \sim 4$–10 observations from \cite{Nakajima23} and \cite{Curti23}, respectively. In the $z \sim 3$ panel, gray symbols represent the sample from \cite{Sanders21}. Red rectangles, shown in all panels, denote the selection criteria for the subsample, which is designed to minimize the sample bias and effects of FMR definition by comparing galaxies within a consistent region of the SFR-$M_*$ plane.
    }
    \label{fig:sfr_sm}
\end{figure*}
We estimate stellar masses and SFRs for all our samples using a consistent methodology. The stellar masses are derived by fitting the SED models to NIRCam photometry, while the SFRs are calculated from emission-line fluxes measured from the NIRSpec spectra.

For the JADES sample, the SED fitting inputs are photometric data from the JADES catalog \citep{Rieke23}, specifically total magnitudes derived from circular aperture photometry on PSF-convolved images at the F444W resolution, with aperture corrections applied. Photometric uncertainties are determined using randomly placed apertures across the images. When available, PRISM spectra are simultaneously fit alongside the photometry. Before fitting, the spectra are corrected for slit losses by normalizing them to match the photometric fluxes, ensuring consistency between spectroscopic and photometric data.
For the CEERS and ERO samples, we adopt the stellar masses from the \citet{Nakajima23} catalog, which were derived using NIRCam photometry. For the DREAMS sample, we perform SED fitting on the publicly available NIRCam photometry. Because DREAMS galaxies are gravitationally lensed, the resulting stellar masses and SFRs are corrected for magnification (see Section \ref{sec:dreams} for details). Our fitting procedure, described below, is identical to that of \citet{Nakajima23}, ensuring a consistent analysis.

We perform SED fitting using the Bayesian inference code Prospector \citep{Johnson21}, following the methodology described in \cite{Harikane23a} and \cite{Nakajima23}. Model spectra are generated with the Flexible Stellar Population Synthesis \citep[FSPS;][]{Conroy09,Conroy10} package, incorporating MIST isochrones \cite{Choi16}. We assume a \cite{Chabrier03} initial mass function (IMF), the \cite{Calzetti00} dust attenuation law, and IGM absorption following \cite{Madau95}. 
A flexible, nonparametric star formation history is adopted, where the first bin spans 0–10 Myr and the remaining four are evenly spaced in log time from 10 Myr to a lookback time corresponding to $z = 30$, with constant SFR assumed within each bin.
In the fitting procedure, we vary the total stellar mass, V-band dust attenuation $\tau_V$, and the star formation history as free parameters, while fixing the stellar metallicity at $Z = 0.2 \, Z_\odot$. We assume a continuity prior for the star formation history and flat priors over the following ranges: $0 < \tau_V < 2$ and $6 < \log(M_*/M_\odot) < 12$. The posterior distributions are sampled using Markov Chain Monte Carlo (MCMC) sampling with the EMCEE algorithm \citep{Foreman-Mackey13}, based on the minimum $\chi^2$.

We derive SFRs for all galaxies from the dust-corrected H$\beta$ emission-line luminosities, following the same procedure as \citet{Nakajima23}. The dust extinction for each galaxy is individually estimated from the Balmer decrement (H$\alpha$/H$\beta$ or H$\gamma$/H$\beta$). We assume the \cite{Gordon03} attenuation law and Case B recombination with an electron temperature of $T_e =$ 17,500 K. The choice of Balmer line is based on its availability and signal-to-noise ratio (S/N). To derive the color excess, $E(B-V)$, we use the line with the higher S/N between H$\alpha$ (if covered) and H$\gamma$, provided its S/N is greater than 3. For galaxies where neither H$\alpha$ nor H$\gamma$ meets this S/N threshold, we apply a correction using the average $E(B-V)$ derived from the stacked spectrum from the corresponding redshift bin (Figure \ref{fig:props} (d)). After correcting the H$\beta$ flux for extinction, we calculate SFRs using the \cite{Kennicutt98} relation, adjusted for a \cite{Chabrier03} IMF via the conversion factor from \cite{Madau14}. Figure \ref{fig:sfr_sm} summarizes the distributions of stellar mass and SFR obtained for our sample.

%
\subsection{Stacking Analysis}
\label{sec:stacking}
\begin{figure*}[htb]
    \centering
    \includegraphics[width=\linewidth]
        {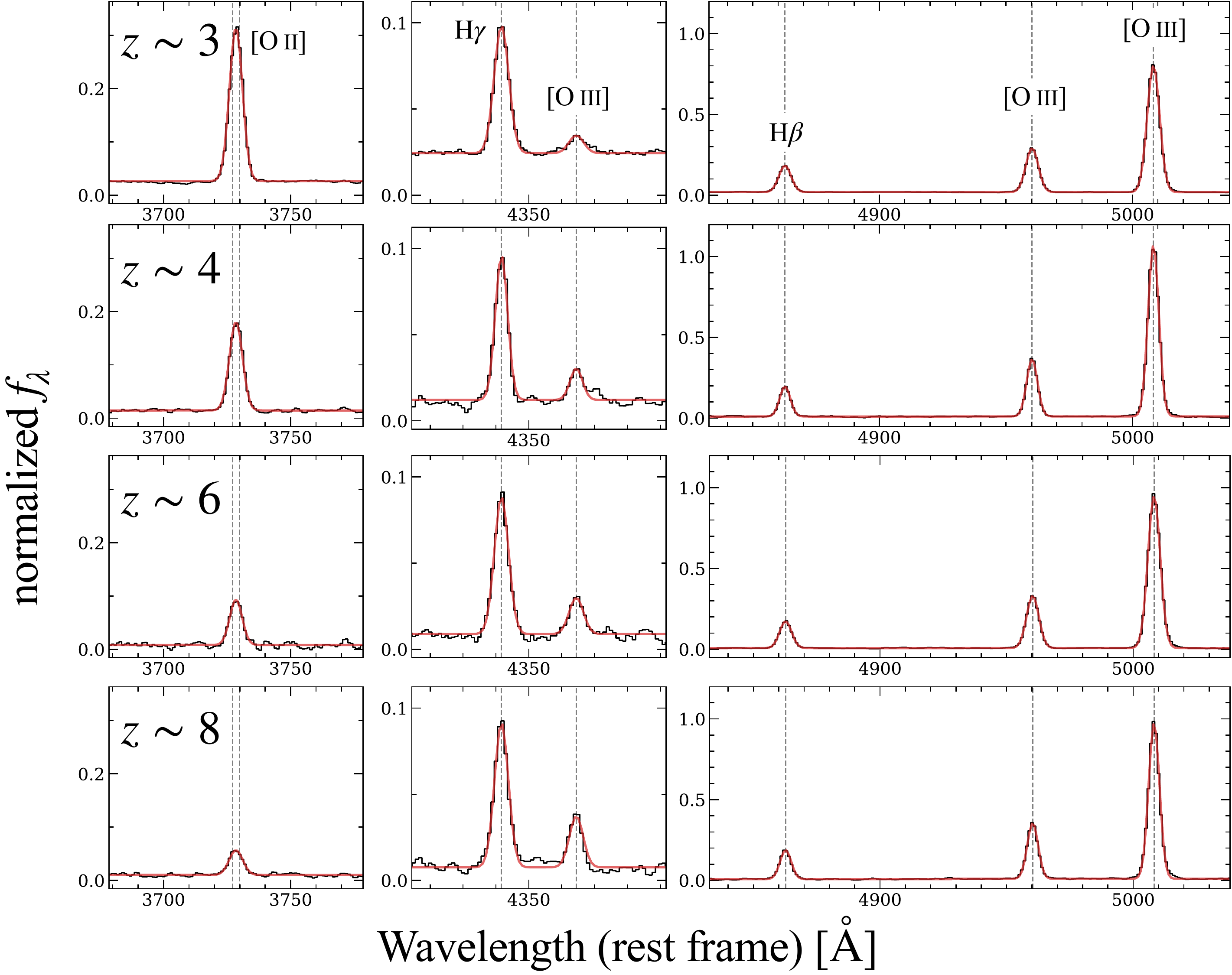}
    \caption{
    Stacked spectra used in this study, shown for four redshift bins ($z \sim 3$, 4, 6, and 8; from top to bottom) and three wavelength ranges (from left to right):
    \OII$\lambda$3727; \Hg\ and \OIIIlam{4363}; and \Hb, \OIIIlam{4959}, and \OIIIlam{5007}.
    Gray vertical lines mark the expected positions of the relevant emission lines.
    Black curves show the stacked spectra normalized by the \Hb\ flux of each galaxy before stacking.  
    Red curves indicate the Gaussian fits used for emission-line flux measurements.
    }
    \label{fig:stack_spectra}
\end{figure*}
\begin{figure*}[htb]
    \centering
    \includegraphics[width=\linewidth]
        {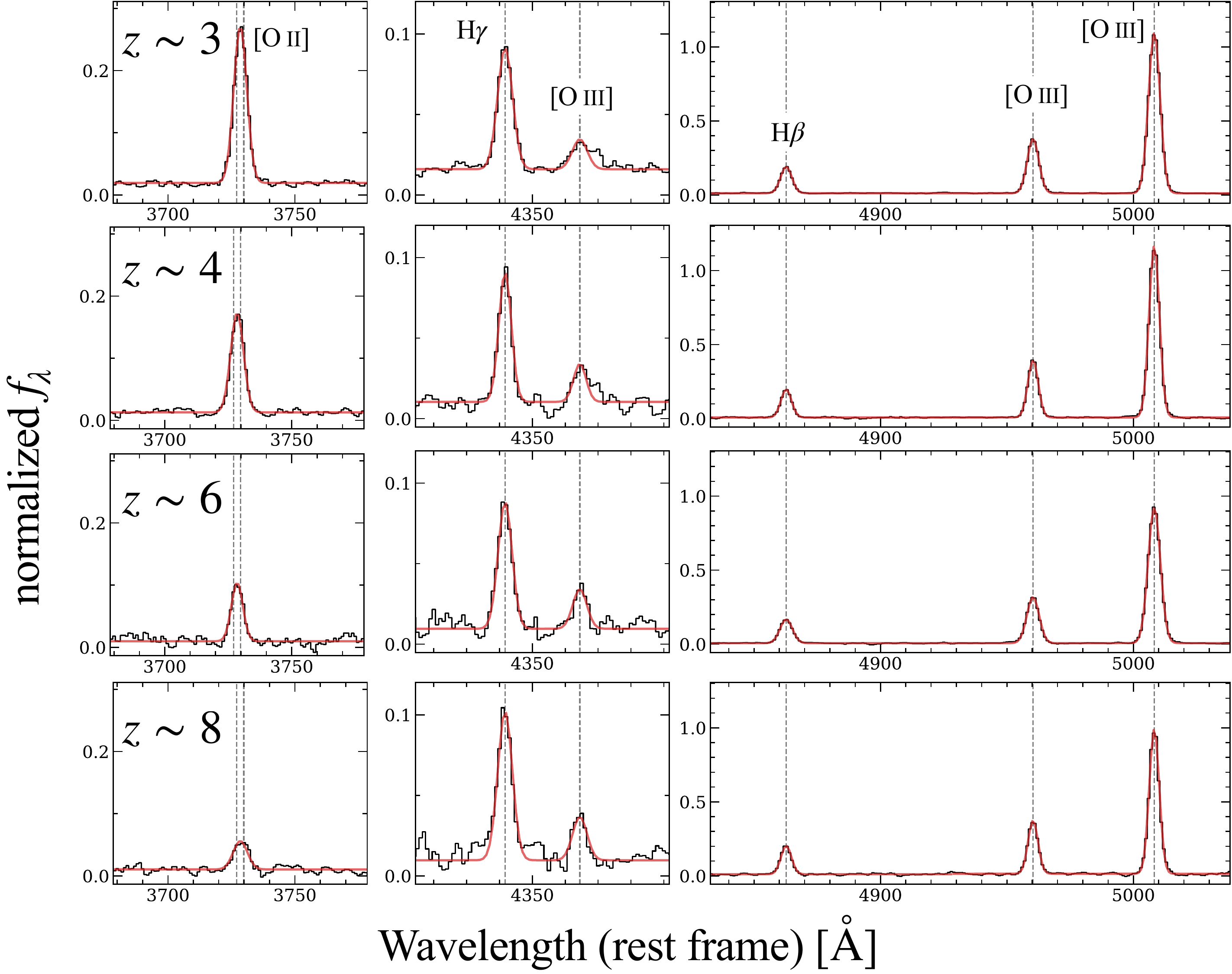}
    \caption{
    Same as Figure \ref{fig:stack_spectra}, but for sample in a fixed $M_*$-SFR subsample.
    }
    \label{fig:stack_spectra_bin}
\end{figure*}
We stack the medium grating spectra of our sample in redshift bins to improve the signal-to-noise ratio. This allows us to derive an average metallicity for each bin, based on the assumption that galaxies within it share broadly similar physical properties, particularly metallicity and emission-line ratios.

We divide the sample into four redshift bins spanning $z=2$ to $z=9.5$, containing $N=$ [122, 60, 49, 18] galaxies with median redshifts of $z=$ [3.1, 4.4, 6.1, 7.9], as summarized in Table \ref{tab:props}.
Before stacking, the individual spectra are shifted to the rest-frame and linearly interpolated onto a common wavelength grid with $\Delta\lambda=1$ \AA\ in linear wavelength space.
This grid spacing is smaller than the typical spectral resolution of the rest-frame, medium-resolution spectra, meaning the process generally involves interpolation. As the emission lines are well-resolved with widths broader than 1 \AA, this interpolation conserves their total flux. We confirmed that using a coarser grid of 2 \AA\ has a negligible impact on our final results.
The each spectrum is normalized by its \Hb\ flux. We then calculate the weighted mean spectrum by applying weights given by the inverse of the S/N of the \Hb\ line flux for each galaxy. 
To reduce the impact of outliers, we apply a sigma clipping procedure at each wavelength point, removing data points deviating by more than three standard deviations from the median. When bad pixels coincided with the positions of emission lines, those pixels are masked before stacking. 

We measure the emission-line fluxes and ratios of the stacked spectra using the same procedures as for individual spectra as described in section \ref{sec:flux_measurement}. 
We correct for dust extinction in the stacked spectra using the Balmer decrement. For the $z \sim 3, 4,$ and $6$ stacks, we utilize the H$\alpha$/H$\beta$ ratio. Note that the H$\alpha$ emission in these stacks is derived from the subset of galaxies ($>80\%$ of the sample) where the line falls within detector coverage, with no contribution from sources located in detector gaps or at $z > 6.9$. For the $z \sim 8$ stack, where H$\alpha$ is redshifted out of the NIRSpec range, we use the H$\gamma$/H$\beta$ ratio. We adopt the \citet{Gordon03} attenuation law and assume Case B recombination with an electron temperature of $T_e = 17,500$ K.
The resulting stacked spectra are shown in Figure \ref{fig:stack_spectra}.

In addition to redshift binning, we also create composite spectra within each redshift bin by constructing the fixed $M_*$–SFR subsample, 
defined as rectangular regions in the $M_*$–SFR plane (Figure \ref{fig:sfr_sm}). This subsampling aims to reduce the dependence on the choice of FMR formulation and to mitigate potential biases related to sample selection and galaxy intrinsic properties. The numbers of galaxies in these subsamples are $N=$ [38, 34, 22, 7], with median redshifts of $z=$ [3.3, 4.5, 6.2, 7.7], also listed in Table \ref{tab:props}. These composite spectra are generated and analyzed following the same procedure as above, and are presented in Figure \ref{fig:stack_spectra_bin}.
\subsection{Uncertainties}
\label{sec:unc}
Uncertainties on derived properties such as emission-line ratios, electron temperatures, and metallicities are estimated using a bootstrap resampling technique. In each realization, we perturb the individual science spectra according to their associated error spectra, resample the galaxy population within each stacking bin with replacement allowing for duplication, and generate stacked spectra following the procedure described above. Emission-line ratios and metallicities are then remeasured from each perturbed stack. We perform 1000 such realizations, and the uncertainties on each measured quantity are estimated as the half-width of the central 68th percentile of the resulting distribution. This approach accounts for both measurement uncertainties and sample variance in the stacked spectra.
\subsection{Photoionization Models}
\label{sec:cloudy}
To interpret the observed emission-line properties, we employ a grid of photoionization models using the \textsc{Cloudy} code \citep[version c23;][]{Ferland17}. This approach provides a physically motivated framework to connect the observed line ratios, particularly \OIII$\lambda5007$/\OII$\lambda$3727 (O32), \OIII$\lambda5007$/\Hb\ (R3), and \OII$\lambda$3727/\Hb\ (R2), with the underlying physical conditions of the ionized gas. Specifically, we use the model grid to estimate the ionization parameter that reproduces the observed O32 ratio at a given metallicity and to test the consistency of our full set of observed line diagnostics.

We construct the model grid by varying two key parameters: the gas-phase metallicity and the ionization parameter. The grid spans four discrete metallicities, $Z=[0.1, 0.2, 0.5, 1.0]\,Z_\odot$, and six ionization parameters, ranging from $\log U=$-3.0 to -0.5 in steps of 0.5 dex. The hydrogen gas density is fixed at $n_\mathrm{e} = 300\,\mathrm{cm^{-3}}$. For the ionizing radiation source, we adopt the stellar continuum from the BPASSv2.3 models for a binary burst population \citep{Eldridge17,Stanway18}. We assume a stellar population with an age of 1 Myr and set the stellar metallicity to match the gas-phase metallicity of each grid point. For the oxygen abundances, we assume a dust depletion factor of 0.6 \citep{Jenkins87}. The calculations are terminated when the electron fraction falls below 0.01, ensuring that the line-emitting region is fully encompassed.
\section{Results} \label{sec:results}
\subsection{Strong Line Ratios}
\label{sec:r_evolution}
\begin{figure}[htb]
    \centering
    \includegraphics[width=\linewidth]
        {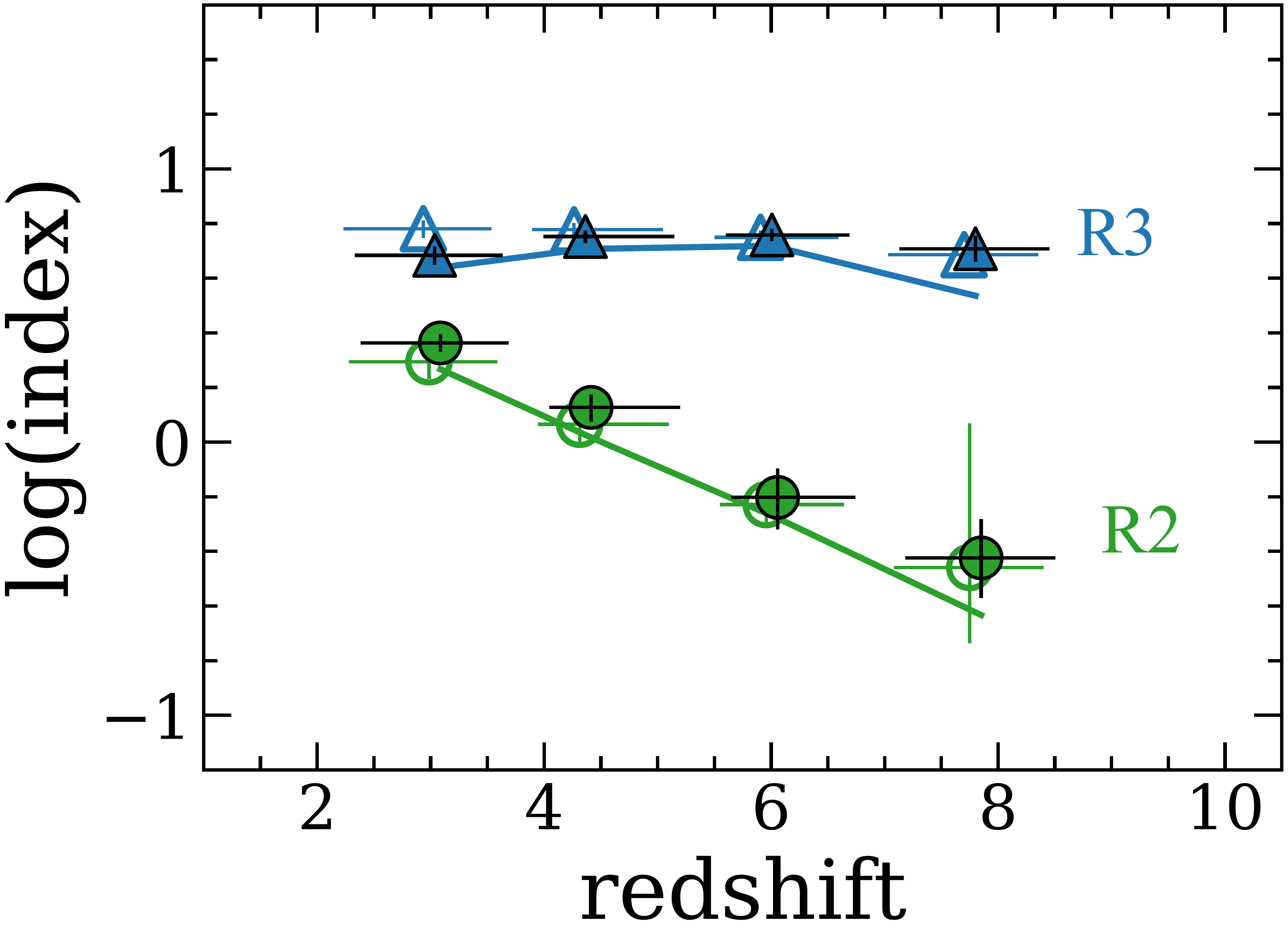}
    \caption{
    Redshift evolution of strong-line indices measured from stacked spectra. The horizontal axis shows redshift ($z \sim 3$, 4, 6, and 8), and the vertical axis shows the logarithmic values of the line ratios: R3 (\OIII$\lambda$5007/\Hb); blue triangles) and R2 (\OII$\lambda$3727/\Hb); green circles).
    Filled symbols represent measurements from the full sample at each redshift, while open symbols correspond to subsamples matched in stellar mass and SFR. The lines indicate predictions from \textsc{Cloudy} photoionization models, calculated using the observed gas-phase metallicity and ionization parameter ($\log U$) derived for each redshift bin from the full sample.
    }
    \label{fig:r_evolution}
\end{figure}

We measure line fluxes for the stacked spectra, including \OII\,$\lambda$3727, \OIIIlam{4363}, \Hb, and \OIII\,$\lambda\lambda$5007,4959. The fitting procedures are same as for individual objects described in Section \ref{sec:flux_measurement}. We then derive line ratios such as R3, R2, and O32 after dust correction.
Figure \ref{fig:r_evolution} presents the evolution of the strong-line indices R3 and R2 as a function of redshift, derived from stacked spectra at $z \sim 3$, 4, 6, and 8. 
The R3 index remains nearly constant across this redshift range, while R2 index decreases, indicating that \OII\ emission becomes relatively weaker at earlier times compared to both \Hb\ and \OIII. These trends persist even when comparing the fixed $M_*$–SFR subsample,
suggesting that they are not driven solely by global galaxy properties but rather reflect conditions in the ISM. We discuss the physical interpretation of this trend with a photoionization model in section \ref{sec:logu}.
\subsection{Direct Method Metallicity}
\label{sec:direct_metallicity}
All stacked spectra exhibit significant detections of the \OIIIlam{4363} auroral line (S/N$>$3), enabling us to determine the gas-phase oxygen abundance using the direct electron temperature ($T_\mathrm{e}$) method. 
Specifically, we measure the S/N of the line to be 4.9, 5.0, 5.0, and 6.7 in our four redshift bins ($z\sim3$, 4, 6, and 8, respectively). The line remains significantly detected even in our fixed $M_*$–SFR subsample, with S/N values of 4.5, 5.0, 3.5, and 5.0.

We first estimate the electron temperature of the $\mathrm{O}^{2+}$ emitting region, $T_\mathrm{e}$(\OIII), from the \OIII\,$(\lambda 4363/\lambda 5007)$ line ratio, assuming an electron density of 300 $\mathrm{cm}^{-3}$. This calculation is performed using the getTemDen routine in the PyNeb package \citep{Luridiana15}. 
The adopted value of 300 $\mathrm{cm}^{-3}$ for the electron density is consistent with that inferred from the \OII\ doublet in the stacked spectra at all redshifts (\ref{sec:density}).
We also confirm that the assumed density has a negligible effect on the derived metallicities, as varying it between 10 and 1000 $\mathrm{cm}^{-3}$ changes log(O/H) by less than 0.03 dex.
The electron temperature of the $\mathrm{O}^{+}$ emitting region, $T_\mathrm{e}$(\OII), is estimated from $T_\mathrm{e}$(\OIII) using the empirical relation provided by \cite{Izotov06}. Ionic abundances are then computed using the getIonAbundance task in PyNeb: $\mathrm{O}^{+}$/$\mathrm{H}^{+}$ from the \OII\,$\lambda\lambda$3727 to \Hb\ ratio and $T_\mathrm{e}$(\OII), and $\mathrm{O}^{2+}$/$\mathrm{H}^{+}$ from the \OIII\,$\lambda\lambda$4959,5007 to \Hb\ ratio and $T_\mathrm{e}$(\OIII). We do not include contributions from higher ionization states (e.g., $\mathrm{O}^{3+}$/$\mathrm{H}^{+}$), following the approximation adopted in \cite{Izotov06}, and we find no evidence for He{\sc ii}$\lambda$4686 emission in the stacked spectra. 
The derived $T_\mathrm{e}$(\OIII) values and oxygen abundances are summarized in Table \ref{tab:props} and displayed in Figure \ref{fig:props} (a) and (b), respectively.
\subsection{Comparing Direct-metallicity Results to Empirical Strong-line Calibrations}
\label{sec:calibratios}
\begin{figure*}[htb]
    \centering
    \includegraphics[width=\linewidth]
        {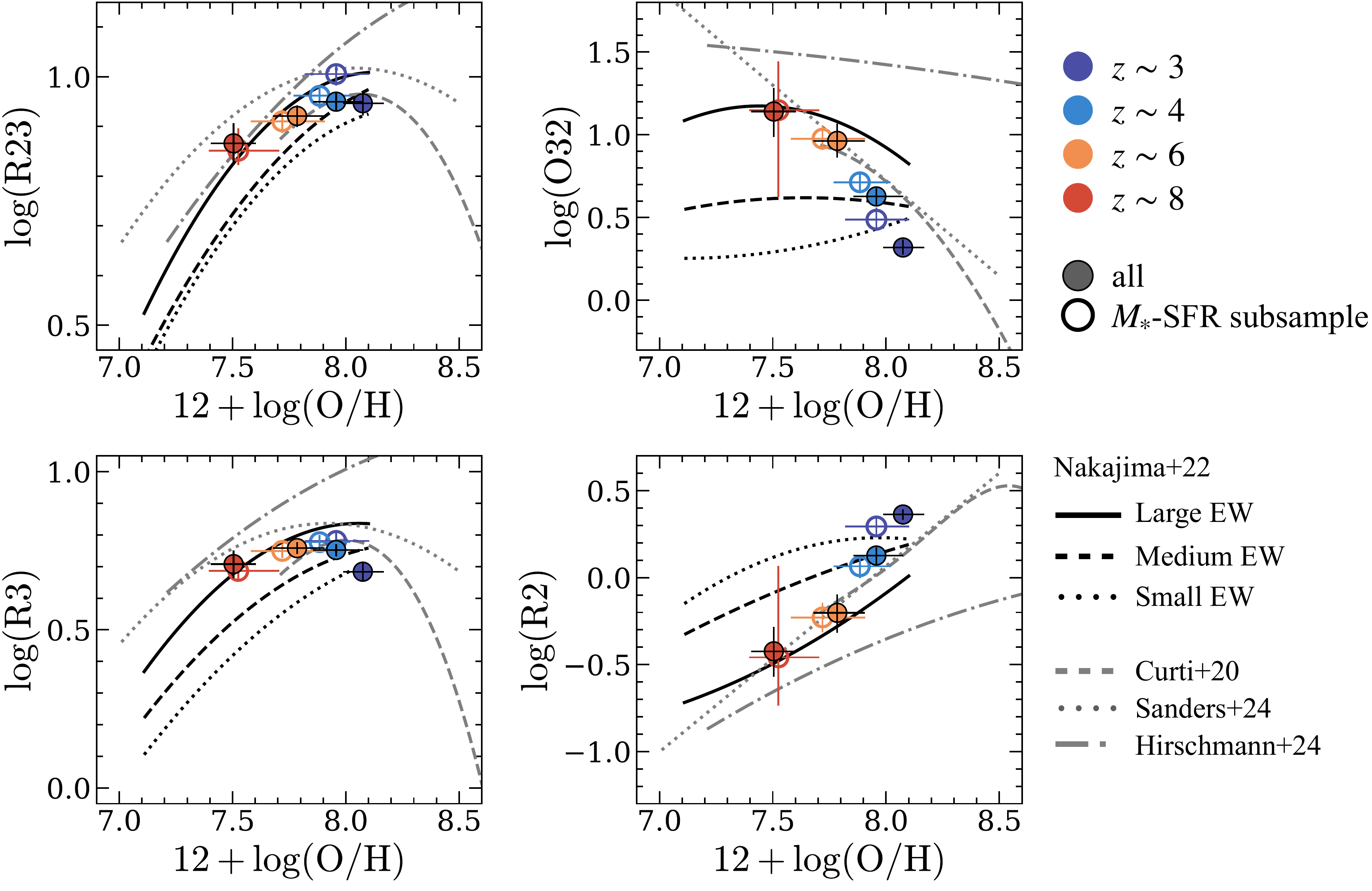}
    \caption{
    Comparison between strong-line ratios and metallicities derived from the direct method. 
    Each panel shows a different emission-line diagnostic: top-left shows R23 $=$ \OIII$\lambda\lambda4959,5007 +$\OII$\lambda$3727)/\Hb,
    top-right shows O32 $=$ \OIII$\lambda5007$/\OII$\lambda$3727, 
    bottom-left shows R3 $=$ \OIII$\lambda5007$/\Hb,
    and bottom-right shows R2 $=$ \OII$\lambda$3727/\Hb.
    Colored data points represent results from stacked spectra at different redshifts: $z \sim 3$ (dark blue), $z \sim 4$ (light blue), $z \sim 6$ (orange), and $z \sim 8$ (red). Filled symbols show the full sample at each redshift, while open symbols indicate sub-samples binned by stellar mass and SFR. Lines show metallicity calibrations from the literature. Black solid, dashed, and dotted lines correspond to the empirical relations from \cite{Nakajima22b} for local galaxies with large (EW(\Hb) $>$ 200 \AA), medium (100 $<$ EW(\Hb) $<$ 200 \AA), and small (EW(\Hb) $<$ 100 \AA) \Hb\ equivalent widths, respectively. The gray dashed line shows the calibration from \cite{Curti20} based on local galaxies. The gray dotted line represents the relation derived from JWST galaxies at $z = 2$--9 by \cite{Sanders24}, and the gray dot-dashed line shows the relation predicted by cosmological simulations from \cite{Hirschmann23}.
    }
    \label{fig:calib}
\end{figure*}
Many previous studies of high-redshift galaxies have relied on empirical metallicity indicators based on strong emission lines. To enable comparisons with these earlier works and to evaluate the applicability of strong-line calibrations to individual galaxies in the early universe, metallicities are also derived using strong-line ratios measured from the stacked spectra. Four widely used diagnostics, (\OIII$\lambda\lambda4959,5007 +$\OII$\lambda$3727)/\Hb\ (R23), O32, R3, and R2, are calculated in each redshift bin. These ratios are then compared to the oxygen abundances obtained from the direct method to assess the validity of existing empirical calibrations. 

Figure \ref{fig:calib} shows the resulting relationships between each strong-line ratio and the metallicity derived via the direct method, along with comparison curves from the literature.
Empirical relations from \cite{Nakajima22b}, \cite{Curti20}, \cite{Sanders24}, and \cite{Hirschmann23} are included in the figure. The relations from \cite{Nakajima22b} and \cite{Curti20} are based on local galaxies, while \cite{Sanders24} derived their relations using JWST observations of galaxies at $z = 2$–9. \cite{Hirschmann23} used a cosmological simulation to establish their calibration. \cite{Nakajima22b} also introduced a method to account for the ionization state of the gas using H$\beta$ equivalent width (EW) as a proxy, motivated by the known correlation between EW(H$\beta$) and ionization-sensitive ratios such as O32. Their calibrations are provided for different EW(H$\beta$) bins: EW $<$ 100 \AA, 100 \AA $<$ EW $<$ 200 \AA, and EW $>$ 200 \AA.

For each redshift bin in this study, the average EW(H$\beta$) of the stacked spectra is measured to be 53 \AA, 100 \AA, 140 \AA, and 140 \AA\ at $z \sim 3$, 4, 6, and 8, respectively. Following the procedure described in \cite{Nakajima22b}, the O32 ratio is first used to determine the appropriate metallicity branch (high or low), and the metallicity is then estimated from the R23 ratio using the corresponding relation. 
At $z \sim 3$, 4, and 6, the strong-line-based metallicities agree well with the results from the direct method. At $z \sim 8$, however, the strong-line calibrations predict systematically higher metallicities than the direct method.  This trend is also reported in previous studies (\citep[e.g.,][]{Nakajima23,Sanders24}), which pointed out that the dependence on EW(H$\beta$) becomes weaker for galaxies at very high redshift. When using the EW $>$ 200 \AA\ calibration, the metallicities and O32 values become more consistent with those derived from the direct method at $z \sim 8$.

Overall, none of the existing calibrations consistently reproduces the observed relationships between strong-line ratios and direct-method metallicities for all four diagnostics across all redshifts. Caution is therefore needed when applying these calibrations to interpret the metallicities of high-redshift galaxies.
\subsection{Interpreting the Evolution of Strong-line Ratios with Photoionization Models}
\label{sec:logu}
\begin{figure}[htb]
    \centering
    \includegraphics[width=\linewidth]
        {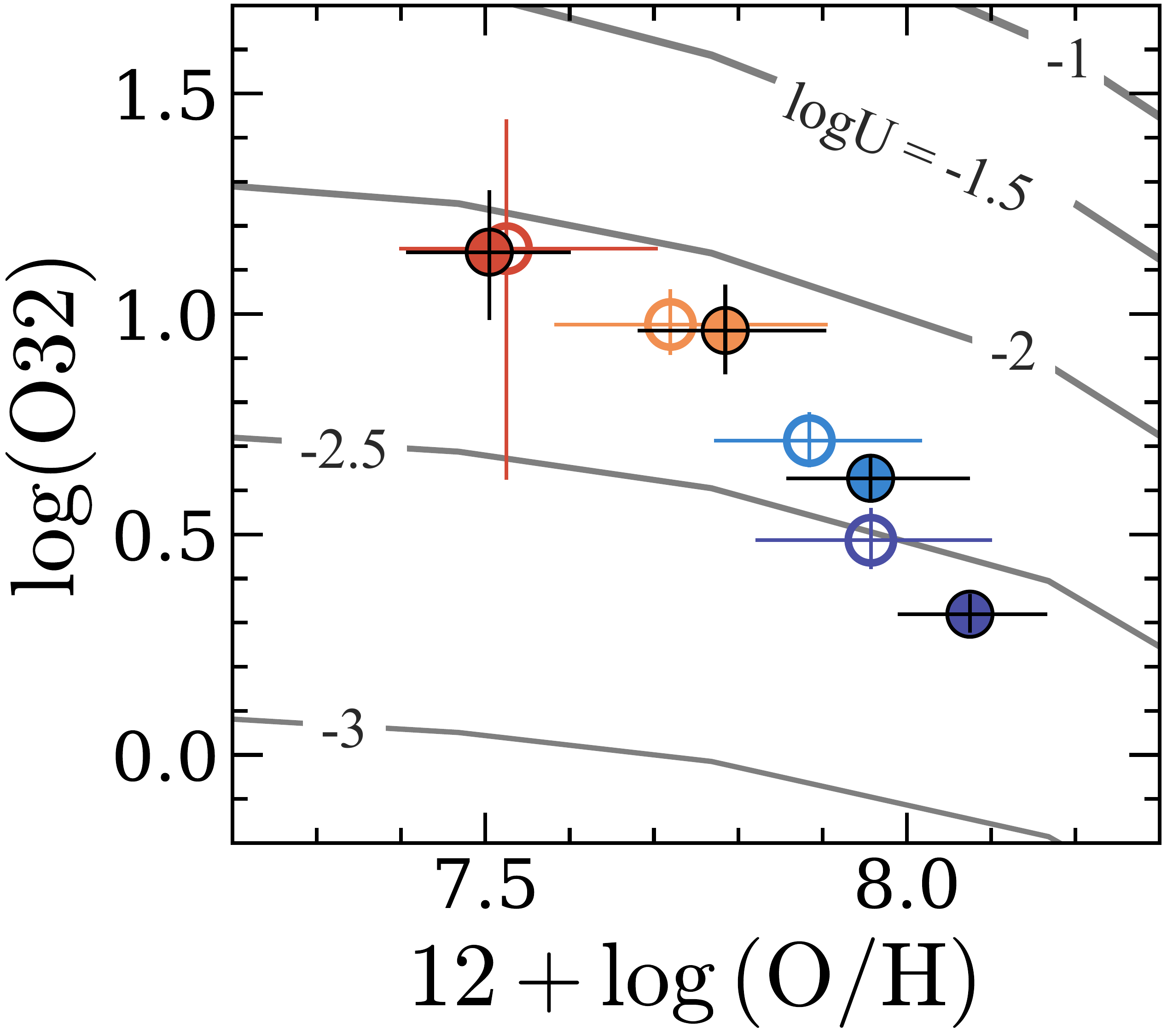}
    \caption{
    Comparison of observed O32 indices with \textsc{Cloudy} photoionization model predictions as a function of gas-phase metallicity. The gray curves show model predictions at different ionization parameters log $U$, ranging from $-3$ (thinnest) to -1.0 (thickest), computed for metallicities $Z = 0.1$, 0.2, 0.5, and 1 $Z_\odot$. Colored circles indicate observed O32 values at $z \sim 3$, 4, 6, and 8, where the colors correspond to redshift as in Figure \ref{fig:calib}. Filled symbols represent the full sample at each redshift, and open symbols correspond to fixed $M_*$-SFR subsample. By comparing the observed data points to the model grid, we estimate the ionization parameter $\log U$ for each redshift bin.
    }
    \label{fig:cloudy}
\end{figure}
\begin{figure}[htb]
    \centering
    \includegraphics[width=\linewidth]
        {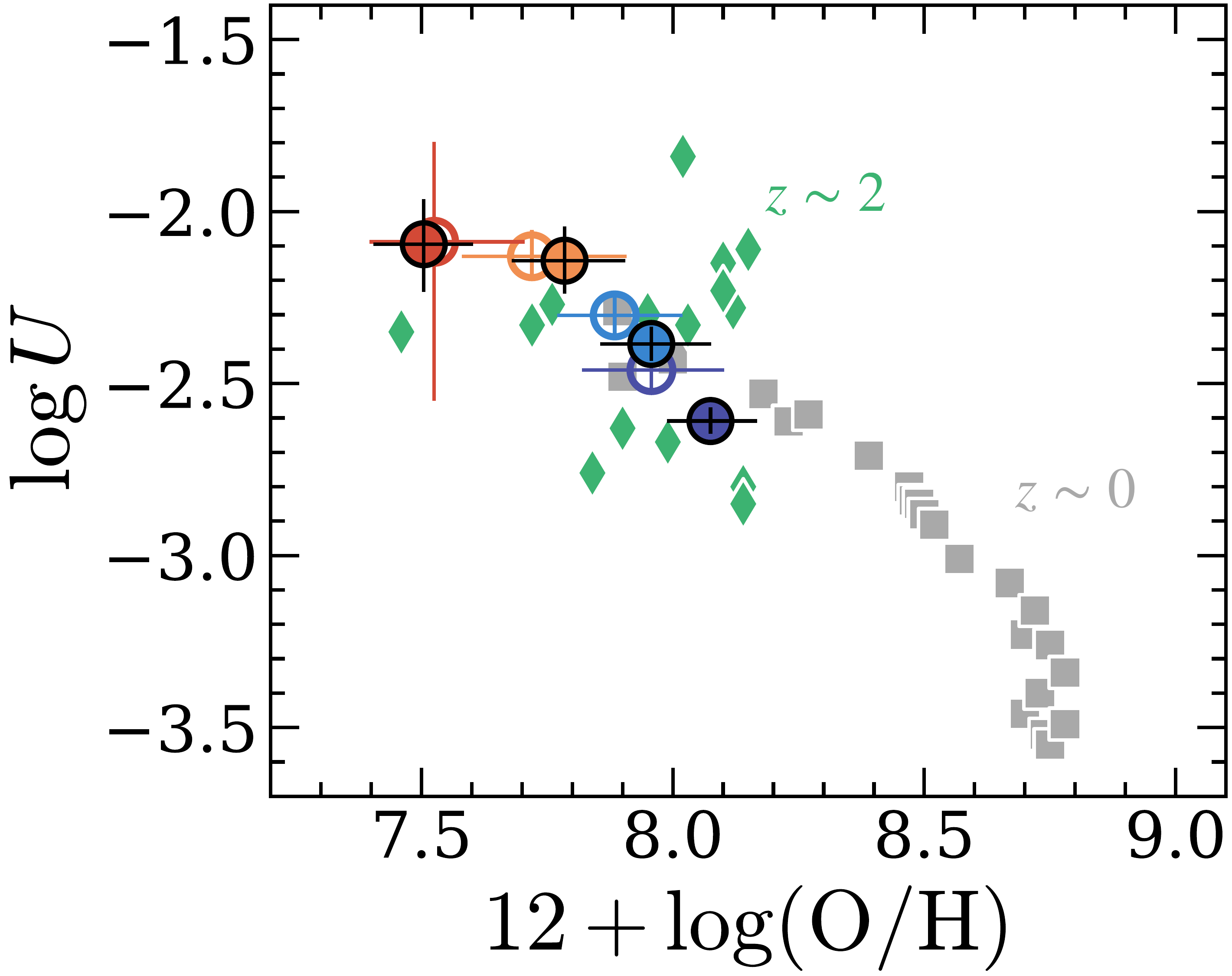}
    \caption{
    Relationship between ionization parameter and gas-phase metallicity.
    The horizontal axis shows gas-phase metallicity derived using the direct $T_e$ method, while the vertical axis shows the ionization parameter $\log U$, estimated via photoionization modeling.
    Colored circles represent stacked galaxy samples at redshifts $z\sim3$, 4, 6, and 8. Filled circles indicate stacks of the entire sample at each redshift, while open circles show stacks of the fixed $M_*$–SFR subsample.
    For comparison, gray squares denote the $z\sim0$ SDSS results \citep{AM13}, and green diamonds show the $z\sim2$ results from the MOSDEF survey \citep{Sanders20}.
    Our sample traces the low-metallicity end of the anti-correlation between metallicity and $\log U$ observed at $z\sim0$ and $z\sim2$.
    }
    \label{fig:logu}
\end{figure}
As shown in Section \ref{sec:r_evolution}, our stacked samples at $z \sim 3$, 4, 6, and 8 show that the R3 index remains roughly constant with redshift, whereas R2 decreases and O32 increases (Figure \ref{fig:r_evolution}). To assess whether these changes can be explained by variations in metallicity and ionization parameters, we employ photoionization models constructed using the \textsc{Cloudy} code, as described in Section \ref{sec:cloudy}. 
 
We first estimate the ionization parameter $\log U$ by computing O32 on a metallicity–ionization parameter grid generated with the photoionization model described in Section \ref{sec:cloudy} (Figure \ref{fig:cloudy}). Interpolating this grid with the observed O32 and metallicity values yields ionization parameters for each redshift bin. The resulting values of $\log U$ increase from approximately $-2.6$ at $z \sim 3$ to $-2.1$ at $z \sim 8$ (Figure \ref{fig:props} (c)).

We present the relation between the ionization parameter and metallicity in Figure \ref{fig:logu}.
Previous studies have established an empirical anti-correlation, observed up to $z \sim 2$–3 \citep{Sanders20}, in which galaxies with lower metallicities tend to have higher ionization parameters.
Our sample at $z > 2$--10 lies at the low-metallicity, high-ionization end of this relation.
This finding indicates that the elevated ionization parameters in high-redshift galaxies are likely a consequence of their low metallicities, rather than evidence for an intrinsic evolution of $\log U$ with redshift.

We next test if our derived physical properties can self-consistently explain the observed trends. Using the estimated metallicities and ionization parameters for each redshift bin as inputs to our photoionization models, we predict the corresponding R3 and R2 values. The model predictions, overplotted in Figure \ref{fig:r_evolution}, show good agreement with the observations. This consistency demonstrates that the observed evolution in strong-line ratios is primarily driven by the joint evolution of gas-phase metallicity and ionization parameter.

The near-constancy of R3 arises from a balance between two competing effects associated with the high-redshift environment: a higher ionization parameter and a lower overall oxygen abundance. The elevated ionization parameter increases the fraction of oxygen in its doubly-ionized state (O$^{++}$), which acts to boost the \OIII\ emission. Simultaneously, the lower total oxygen abundance reduces the number of available oxygen ions, which suppresses the emission. These two competing effects largely counteract each other, resulting in a relatively stable R3 across redshift.
In contrast, the decline in R2 is driven by the decreasing abundance of O$^{+}$ ions. This decrease is caused by both a lower total oxygen abundance and the efficient conversion of O$^{+}$ to O$^{++}$ in the increasingly high-ionization conditions at higher redshifts, reaching $\log U\sim-2$ at $z\sim8$.
\subsection{Electron Density}
\label{sec:density}
\begin{figure*}[htb]
    \centering
    \includegraphics[width=\linewidth]
        {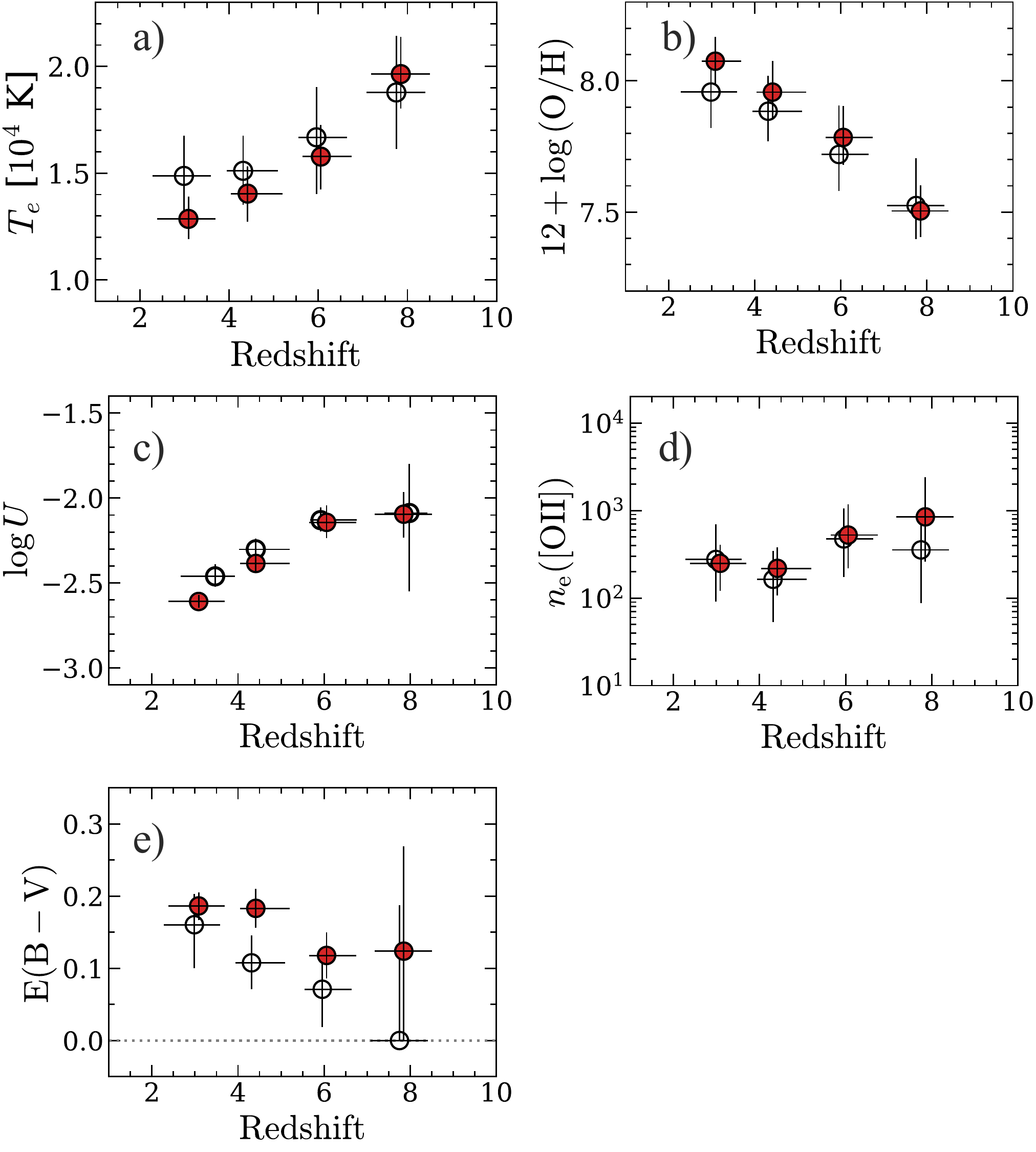}
    \caption{
    Redshift evolution of key physical properties derived from our sample. The panels show: (a) electron temperature, (b) gas-phase metallicity, (c) ionization parameter, (d) electron density, and (e) nebular dust attenuation $E(B-V)$.
    All parameters are estimated from stacked spectra in each redshift bin.
    Filled red circles indicate measurements from the full sample at each redshift, while open black circles correspond to the fixed $M_*$–SFR subsample.
    }
    \label{fig:props}
\end{figure*}
We estimate the electron density of our sample from the \OII$\lambda$3727 doublet and examine its evolution with redshift. 
Although the \OII\ doublet is not fully resolved in the medium-resolution data, we fit the blended profile with a double-Gaussian model, fixing the line widths and central wavelengths. The lower-resolution stacks contain more objects and therefore yield more stable measurements, and we adopt these results as our fiducial values. The derived electron densities are shown in Figure \ref{fig:props} (d), indicating a mild increase in density with redshift.

Given the large uncertainties in the electron density estimates, we adopt a fixed value of $n_{\rm e} = 300\,\mathrm{cm^{-3}}$ in the \textsc{Cloudy} modeling (Section \ref{sec:cloudy}) and in the metallicity measurements (Section \ref{sec:direct_metallicity}). Although the \OII-based measurements allow for densities up to  $n_{\rm e} \sim 1000\,\mathrm{cm^{-3}}$, assuming this higher value in our analyses does not change the results beyond the measurement uncertainties.
\subsection{Mass-Metallicity Relation}
\label{sec:mzr}
\begin{figure}[htb]
    \centering
    \includegraphics[width=\linewidth]
        {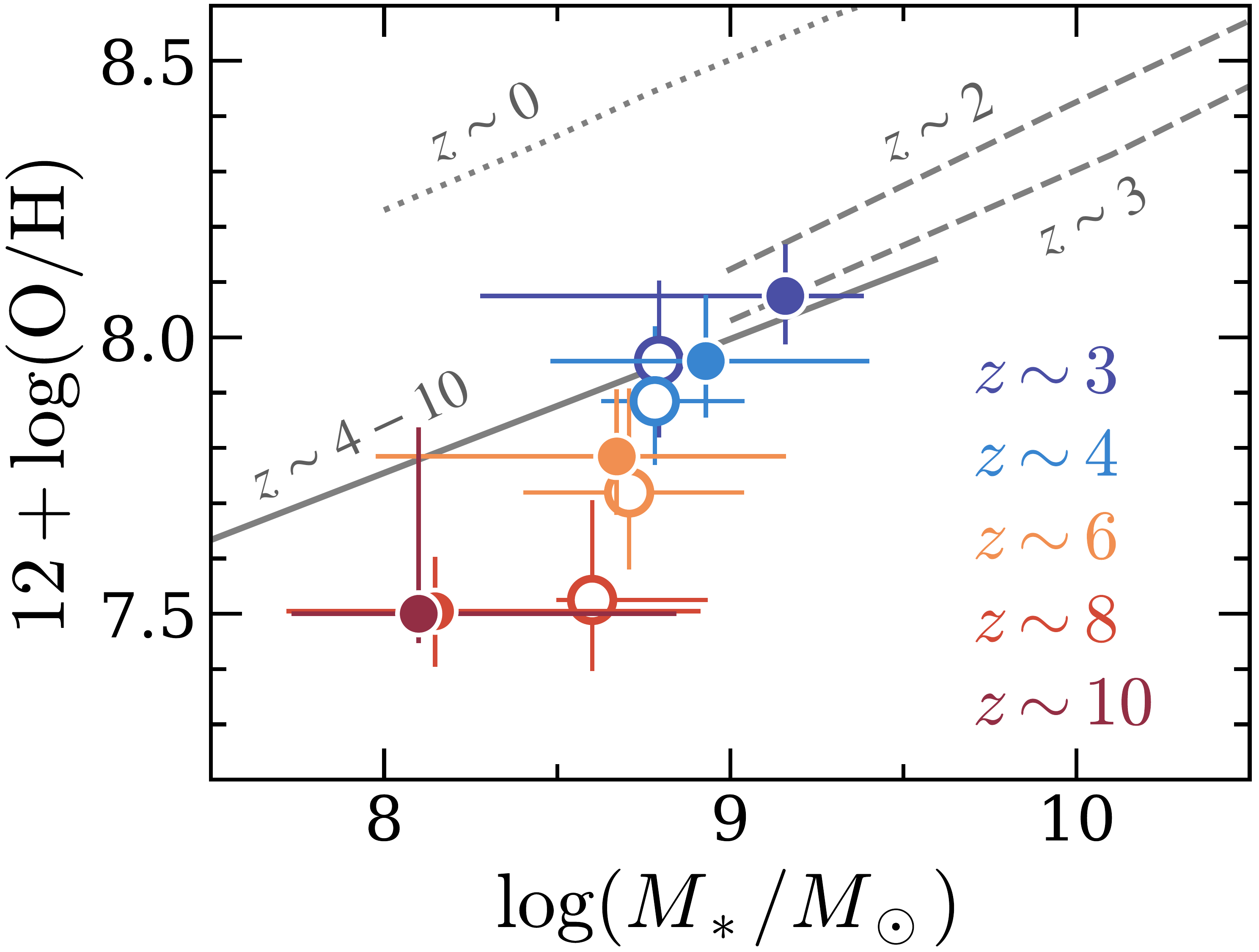}
    \caption{
    Mass–metallicity relation (MZR). 
    Colored data points represent results from stacked spectra at different redshifts: $z \sim 3$ (dark blue), $z \sim 4$ (light blue), $z \sim 6$ (orange), and $z \sim 8$ (red). Filled symbols show the full sample at each redshift, while open symbols indicate sub-samples binned by stellar mass and SFR. Solid and dashed lines show metallicity calibrations from the literature. 
    For $z \sim 2$–8, metallicities are derived from stacked spectra using the direct method, and the error bars show the 68th percentile range obtained via bootstrap resampling. The stellar masses are estimated individually through SED fitting, and the error bars indicate the 68th percentile distribution of the sample at each redshift. At $z \sim 10$, we compile individual metallicity measurements from the literature and show their distribution.
    We also include previous MZR results for comparison: the gray dotted line shows the local relation at $z \sim 0$ \citep{AM13}, the gray dashed lines correspond to $z \sim 2$ and $z \sim 3$ results from \cite{Sanders20}, and the gray solid line represent $z \sim 4$–10 observations from \cite{Nakajima23}.
    Our results suggest a gradual decrease of metallicity at fixed stellar mass towards higher redshifts.
    }
    \label{fig:mzr}
\end{figure}
Figure \ref{fig:mzr} shows the mass–metallicity relation of our sample.
Each data point represents a redshift bin ($z \sim$ 3, 4, 6, 8, 10), with the stellar mass defined as the median of individual galaxy masses within each bin. 
For $z \sim 3-8$, metallicities are derived using the direct method applied to stacked spectra.
At $z \sim 10$, where \OIIIlam{5007} is redshifted out of the NIRSpec coverage and the direct method cannot be applied consistently, we use the median of metallicities from individual galaxies derived using either the direct or strong-line methods (see also Section \ref{sec:high-z}).

These measurements are broadly consistent with previous studies of the MZR at $z = 4\text{--}10$ \citep{Nakajima23,Curti24}. At a fixed stellar mass, we find a clear trend of chemical evolution. The metallicity decreases monotonically from $z \sim 3$ to $z \sim 8$, with a particularly steep decline emerging at $z \gtrsim 8$.
While the $z \sim 10$ measurement relies on a heterogeneous mix of methods and exhibits large dispersion, the overall trend points to a rapid chemical enrichment phase in the early universe. This significant evolution of the MZR at fixed stellar mass raises the question of whether this trend is driven solely by the higher star formation rates typical of high-redshift galaxies (i.e., within the framework of the Fundamental Metallicity Relation), which we address in the following sections.
\subsection{Fundamental Metallicity Relation}
\label{sec:fmr}
\begin{figure*}[htb]
    \centering
    \includegraphics[width=\linewidth]
        {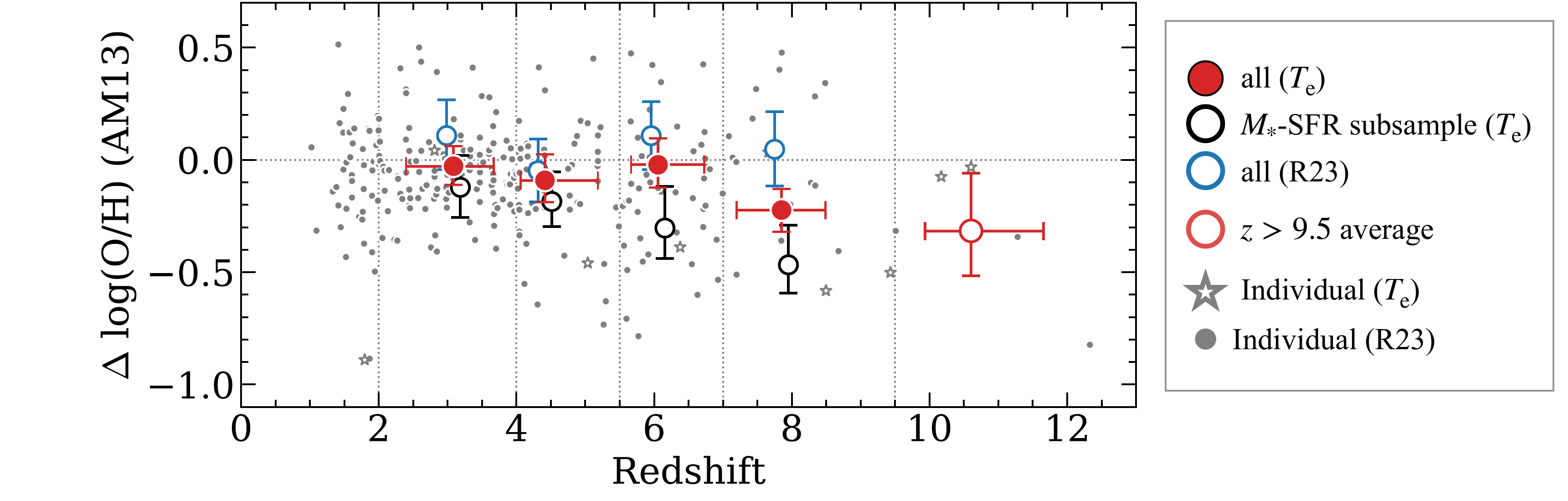}
    \caption{
    Redshift evolution of the deviation from the Fundamental Metallicity Relation (FMR), defined as the difference between the observed metallicity and that predicted by the \cite{AM13} FMR given the stellar mass and SFR.
    Filled red circles indicate our main results at $z \sim 3$, 4, 6, and 8, derived using stacked spectra and the direct method. The open red circle at $z \sim 10$ represents the average of individual measurements from the literature.
    Black red circles show results obtained by stacking galaxies in a fixed $M_*$-SFR range, while blue circles denote metallicities estimated using the R23 strong-line method. 
    Gray symbols show individual galaxies, metallicities derived from the R23 index (dots) or the direct method where possible (stars). 
    A systematic deviation from the local FMR is observed, which becomes most pronounced at $z\gtrsim8$.
    }
    \label{fig:fmr_obs}
\end{figure*}
\begin{figure*}[htb]
    \centering
    \includegraphics[width=\linewidth]
        {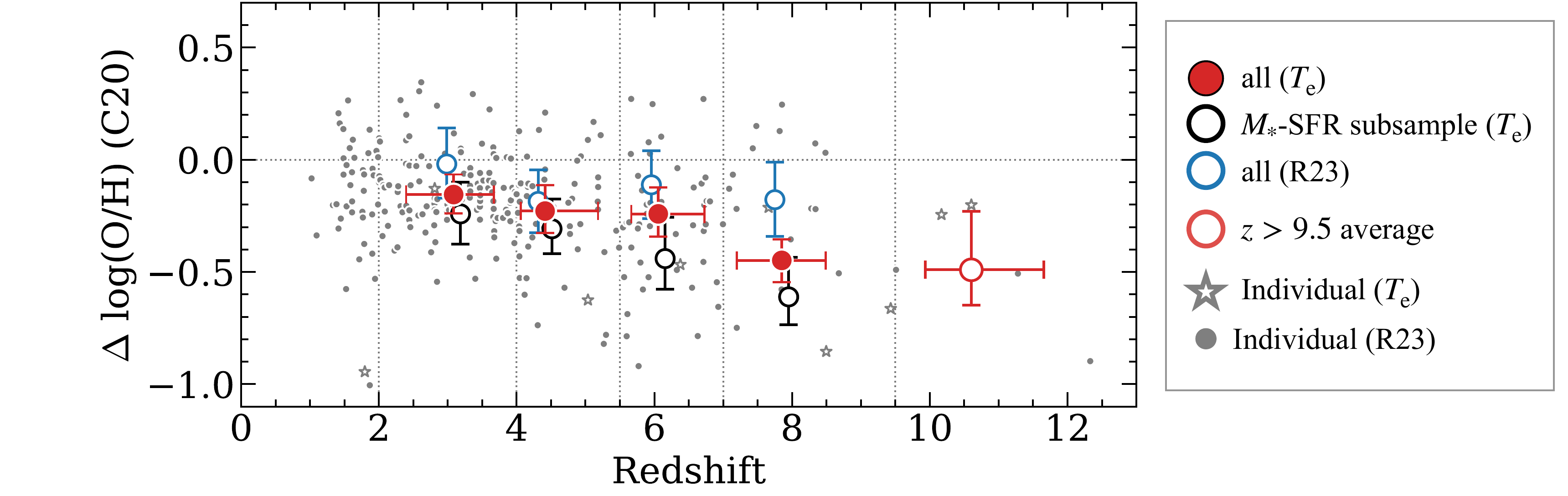}
    \caption{
    Same as the top panel, but for the \cite{Curti20} FMR.
    }
    \label{fig:fmr_obs_C20}
\end{figure*}  
\begin{figure*}[htb]
    \centering
    \includegraphics[width=\linewidth]
        {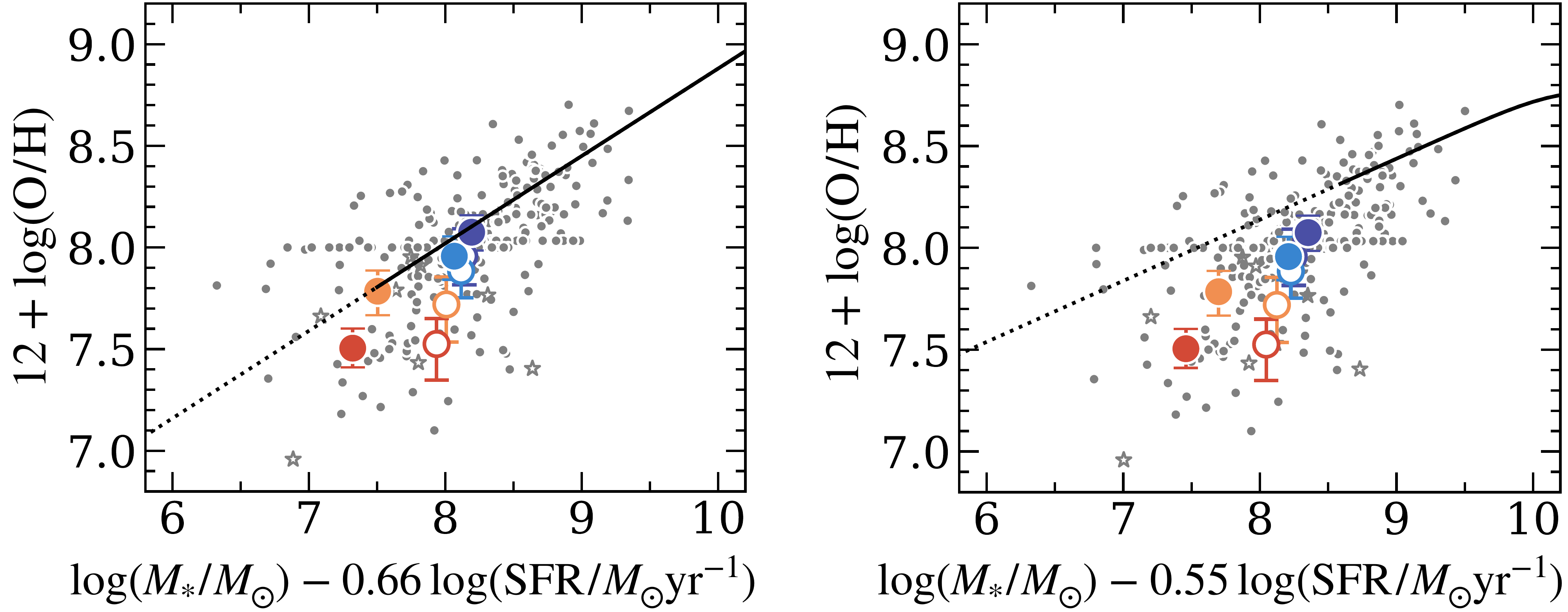}
    \caption{
    Projection of the FMR onto the $\mu_\alpha$--metallicity plane, where $\mu_\alpha = \log M_* - \alpha \log \text{SFR}$.
    The colored symbols represent our stacked results at $z \sim 3$--$8$, as indicated in the legend.
    \textbf{Left:} Projection using $\alpha=0.66$ based on the \cite{AM13} FMR. The black solid curve indicates the range of $\mu_{0.66}$ effectively explored by the \cite{AM13} sample, while the dotted curve represents the extrapolation to lower $\mu_{0.66}$. 
    Data points at $z \lesssim 6$ and all fixed $M_*$--SFR  subsample lie within the explored region. At $z \sim 8$, while the average falls slightly into the extrapolated regime, roughly half of the individual galaxies remain within the explored space.
    \textbf{Right:} Projection using $\alpha=0.55$ based on the \cite{Curti20} FMR. The solid curve shows the constrained range and the dotted curve shows the extrapolation. In this projection, majority of our samples ($z \sim 3$--$8$) reside in the unexplored parameter space, requiring extrapolation of the local relation.
    }
    \label{fig:fmr_alpha}
\end{figure*}
To further investigate this metallicity evolution, we examine the fundamental metallicity relation (FMR), which incorporates both stellar mass and star formation rate. The FMR is a potentially more fundamental relation than the MZR alone, as it is known that metallicity depends on SFR at fixed mass. Moreover, FMR-based comparisons can offer insights into galaxy evolution and help mitigate selection biases due to sSFR.

In Figure \ref{fig:fmr_obs}, we present the difference between the observed metallicities and those predicted by the FMR of \cite{AM13}, 
$\Delta\mathrm{(O/H)} = \mathrm{(O/H)_{obs}} – \mathrm{(O/H)_{pred}}$,
as a function of redshift. 
Here, $\mathrm{(O/H)_{obs}}$ is measured directly from the stacked spectra. To derive $\mathrm{(O/H)_{pred}}$, we calculate the representative stellar mass and SFR for each bin by taking the weighted average of the individual galaxies, using their H$\beta$ signal-to-noise ratios as weights consistent with our stacking procedure. These weighted mean properties are then substituted into the FMR relation.
The redshift values and their uncertainties represent the median and 68th percentile ranges of individual galaxy redshifts. The central values and error bars of $\Delta\mathrm{(O/H)}$ are derived from the same bootstrap procedure described above. The $z \sim 10$ result is based on individual measurements.

Figure \ref{fig:fmr_obs} reveals the evolutionary trend of the FMR offset. For the full sample (red circles), we find that the metallicities are consistent with the predictions of the local FMR up to $z \sim 6$, with $\Delta\mathrm{(O/H)}$ remaining consistent with zero. However, a significant deviation emerges at $z \gtrsim 8$. At these highest redshifts ($z \sim 8$ and $10$), we observe a significant negative offset of $\Delta\mathrm{(O/H)} < -0.2$ dex ($> 1\sigma$). This suggests a potential breakdown of the FMR in the very early universe, indicating that galaxies at $z \gtrsim 8$ are systematically more metal-poor for their stellar mass and SFR than their counterparts.

To assess the robustness and physical dependencies of this result, we show results using the fixed $M_*$–SFR subsample (black symbols).
The metallicities derived from this subsample are consistent with those of the full sample within the uncertainties across all redshift bins.
However, we note that the subsample exhibits a tentative negative offset already starting from $z \sim 3$, appearing to deviate from the local FMR earlier than the full sample trend.
This discrepancy may hint at a mass dependence in the FMR evolution. 
However, investigating this dependence in detail is difficult due to the limited sample size and large uncertainties at the highest redshift bins.

For comparison, Figure \ref{fig:fmr_obs} displays the FMR offset derived using metallicities from the R23 strong-line index on our stacked spectra (blue circles). The R23-based metallicities show no significant deviation from the local FMR even at $z \sim 8$, in stark contrast to the direct-method results.
This suggests that using local strong-line calibrations at high redshift can lead to an overestimation of metallicity (as discussed in \ref{sec:calibratios}), potentially masking the true evolutionary trends. This reinforces the necessity of using the direct method to accurately probe the baryon cycle in the early universe.

Additionally, we plot measurements for individual galaxies as gray dots. Their metallicities are derived using the direct $T_e$ method if the \OIIIlam{4363} line is detected with S/N $>$ 3, and from the R23 index otherwise. 
While these individual measurements exhibit large scatter, their median values in each redshift bin also show a consistent values of $\Delta\mathrm{(O/H)}$ with the primary result obtained from our stacked spectra.

We also explore the impact of the FMR definition by performing the same analysis using the FMR of \cite{Curti20}, as shown in Figure \ref{fig:fmr_obs_C20}. In contrast to the \cite{AM13} case, this version predicts systematically higher metallicities, with $\Delta\mathrm{(O/H)}$ values of $-0.2$ dex already apparent at $z \sim 3$, reaching $-0.4$ dex at $z \sim 8$.

The discrepancy likely stems from the differences in the parameter space covered by the local calibration samples. To demonstrate this, Figure \ref{fig:fmr_alpha} presents the projection of the FMR onto the $\mu_\alpha \equiv \log M_* - \alpha \log \text{SFR}$ plane, where $\alpha$ quantifies the strength of the SFR dependence.
In the left panel using $\alpha=0.66$ from \cite{AM13}, the vast majority ($\sim$93\%) of our sample lies within the parameter space directly constrained by their analysis ($\mu_{0.66} \gtrsim 7.5$). Even for the $z \sim 8$ bin, where the average falls slightly into the unexplored regime, approximately half of the individual galaxies remain within the explored parameter space. Furthermore, the subsamples with fixed $M_*$ and SFR (open circles) consistently fall within the explored region at all redshifts.
Conversely, in the right panel, using $\alpha=0.55$ from \cite{Curti20}, only 34\% of our sample falls within the parameter space directly explored by their analysis ($\mu_{0.55} \gtrsim 8.5$), meaning the relation relies on extrapolation for high-redshift galaxies.
These comparisons confirm that the \cite{AM13} relation is well-suited for our study, as its calibration sample includes the low-mass, actively star-forming galaxies characteristic of the high-redshift universe. Additionally, the strong SFR dependence ($\alpha=0.66$) inherent to the \cite{AM13} formulation is consistent with the high specific star formation rates observed at these redshifts; indeed, \cite{Curti20} report a similarly strong dependence ($\alpha \approx 0.65$) for their high-sSFR population. Therefore, we adopt the \cite{AM13} FMR as the appropriate baseline for assessing metallicity evolution in this study.
\section{Discussion: Possible Physical Origins of the FMR Break}
\label{sec:discussion}
Our findings indicate that high-redshift galaxies systematically deviate from the local FMR, with a significant deviation at $z\sim$8, where galaxies exhibit lower gas-phase metallicities than predicted from their stellar mass and star formation rate.
This implies changes in the physical processes governing the baryon cycle at these early epochs. Here we discuss two plausible scenarios.
\subsection{Enhanced Gas Cycling via Inflows and Outflows}
The FMR is often interpreted as a reflection of an equilibrium state, in which galaxies regulate their star formation and chemical enrichment through a balance of gas inflows, star formation, and outflows
\citep[e.g.,][]{Dave12,Lilly13}. The observed drop in metallicity at high redshifts implies that galaxies at this epoch may not have reached such a self-regulated state. 
In such a scenario, galaxies at high redshifts are likely to undergo repeated cycles of intense star formation bursts followed by strong outflows. The enriched gas can be expelled into the IGM, and subsequent accretion of metal-poor gas from the IGM can lower the metallicity of the ISM. 
A key reason for this could be that the supply of pristine gas via inflows overwhelms the ejection of enriched gas via outflows, thereby breaking the equilibrium.
The cosmological hydrodynamic simulations by \cite{Harada23} illustrate this exact mechanism, showing that the outflow rate remains nearly constant at $z > 5$, whereas the inflow rate increases with redshift as approximately $(1+z)^{2.5}$. For halos with $\log M_{\mathrm{h}}/M_\odot \approx 11$, which correspond to galaxies in our sample with $\log M_{\ast}/M_\odot \approx 9$, the inflow rate starts to exceed the outflow rate at $z \sim 5–6$. 
This transition directly enhances the dilution of metals in the ISM through the accretion of low-metallicity gas. 
\subsection{Understanding the Evolution of the FMR with Baryon Cycling Models}
\begin{figure*}[htb]
    \centering
    \includegraphics[width=\linewidth]
        {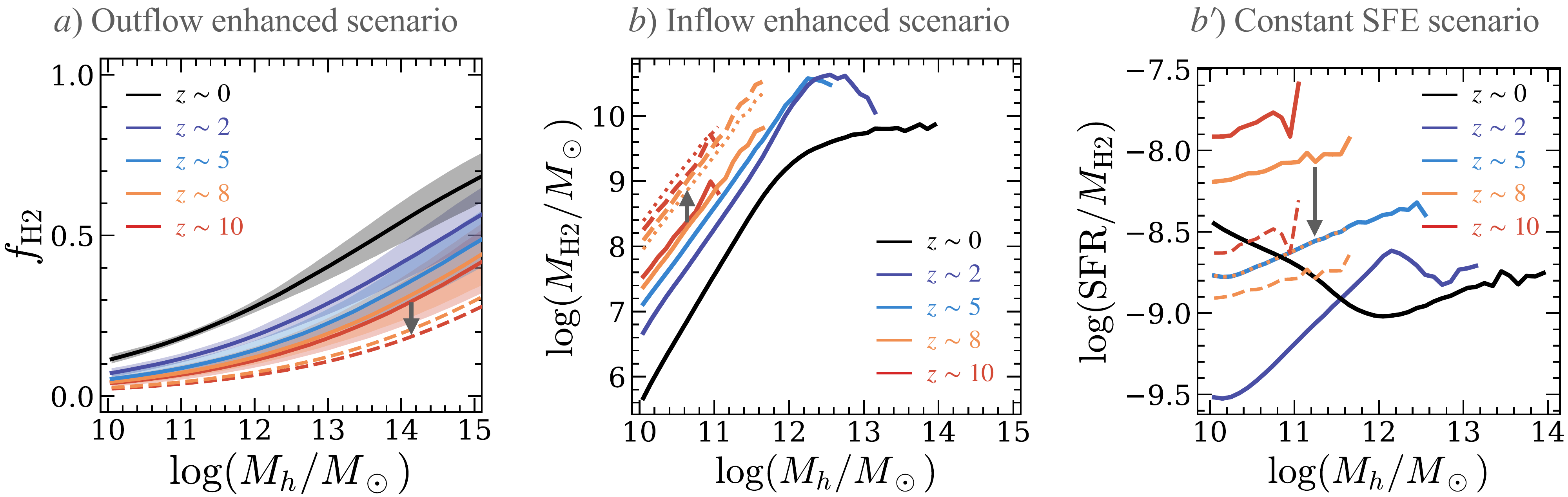}
    \caption{
    \textbf{Left}:
    The $\mathrm{H_2}$ distribution fraction ($f_{\mathrm{H2}}$) predicted by the \textsc{ChemicalUniverseMachine} as a function of halo mass. Different colors correspond to redshifts $z \sim 0, 3, 4, 6,$ and $8$. Solid lines show the standard extrapolation and shaded regions represent the 68th percentile distributions. Dashed lines represent Scenario ($a$), which assumes a lower $f_{\mathrm{H2}}$ at the fixed halo mass for $z > 5$.
    \textbf{Middle}:
    The molecular gas mass ($M_{\mathrm{H2}}$) as a function of halo mass. Colors are the same as in the left panel. Solid lines show the standard extrapolation. Dashed lines represent Scenario ($b$), which assumes a higher $M_{\mathrm{H2}}$ at the same halo mass. Dotted lines indicate Scenario ($b^\prime$) for reference, which shows gas masses nearly identical to Scenario ($b$). 
    \textbf{Right}:
    The star formation efficiency ($\mathrm{SFE} = \mathrm{SFR}/M_{\mathrm{H2}}$) as a function of halo mass. Colors and line styles are the same as in the middle panel. Dotted lines represent Scenario ($b^\prime$), where the SFE is fixed at the $z=5$ value for $z > 5$. Dashed lines show Scenario ($b$) for reference, which yields SFE values comparable to those at $z \sim 0\text{--}5$.
    }\label{fig:scenarios}
\end{figure*}
\begin{figure*}[htb]
    \centering
    \includegraphics[width=0.7\linewidth]
        {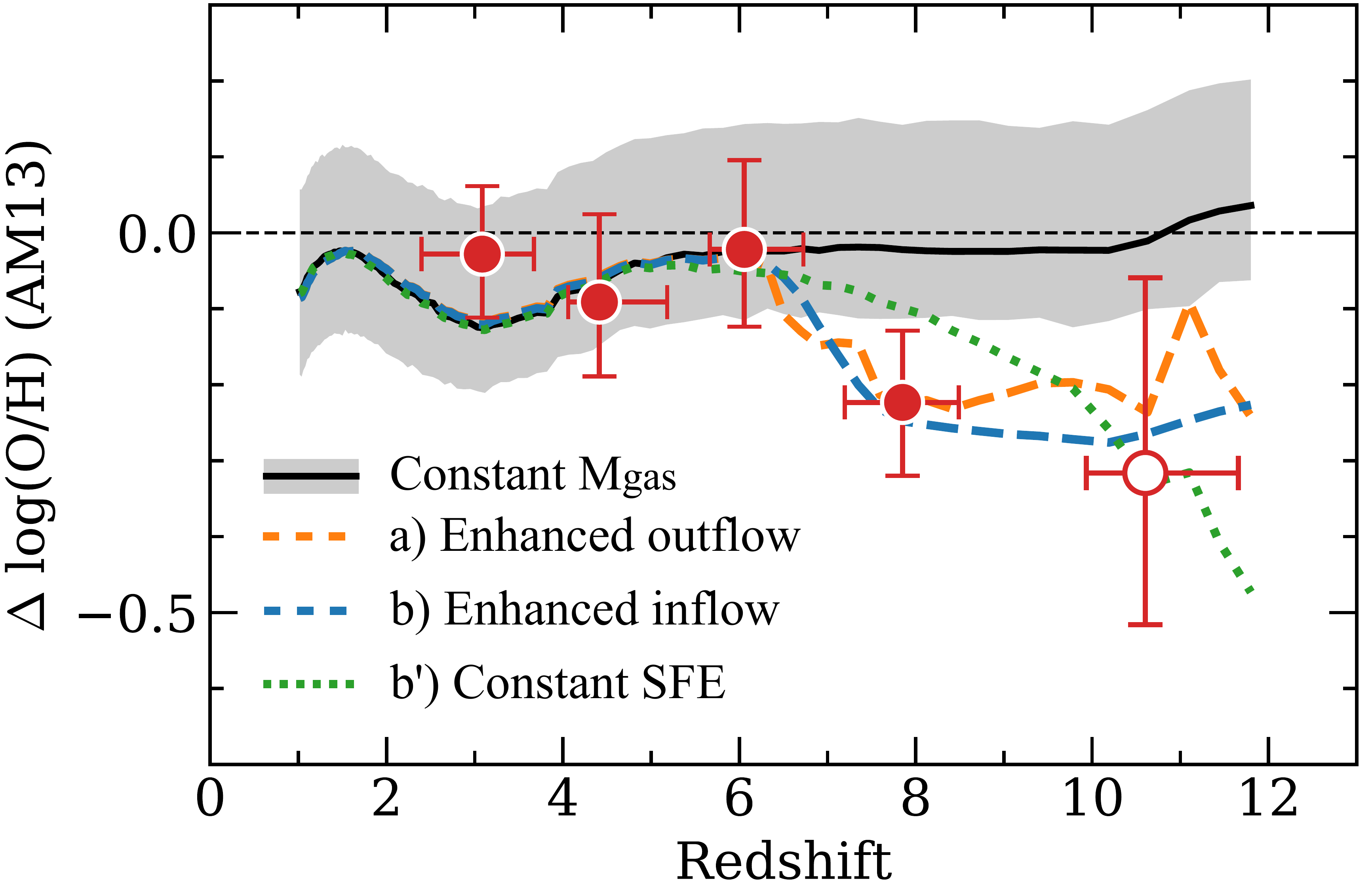}
    \caption{
    FMR evolution shown with predictions from the \textsc{ChemicalUniverseMachine} model \citep{Nishigaki25}.
    Filled and open red circles are the same as in Figure \ref{fig:fmr_obs}: filled circles represent our results derived from stacked spectra using the direct method at $z \sim 3-8$, while the open circle at $z \sim 10$ shows the average metallicity of individual galaxies.
    The curves and shaded regions represent the median predictions and the 68th percentile ranges, respectively, for halos with $\mathrm{SFR} = 10\,M_\odot\,\mathrm{yr^{-1}}$.
    The black line shows standard extrapolation where the molecular gas mass remains roughly constant at fixed halo mass. This scenario predicts no significant deviation from the FMR, failing to reproduce the observed drop at $z > 8$.
    The colored lines represent different physical scenarios at $z > 5$.
    The orange dashed line shows the Enhanced outflow Scenario (Scenario $a$), where the metal retention fraction ($f_{\mathrm{H2}}$) is reduced to model efficient metal removal. The blue dashed line shows the Enhanced inflow Scenario (Scenario $b$), where the molecular gas mass ($M_{\mathrm{H2}}$) is explicitly increased to model strong dilution.
    The green dotted line shows the Constant SFE Scenario (Scenario $b^\prime$), where the star formation efficiency is kept constant, implicitly leading to larger gas reservoirs similar to the Scenario ($b$).
    }
    \label{fig:fmr_um}
\end{figure*}
\begin{figure*}[htb]
    \centering
    \includegraphics[width=\linewidth]
        {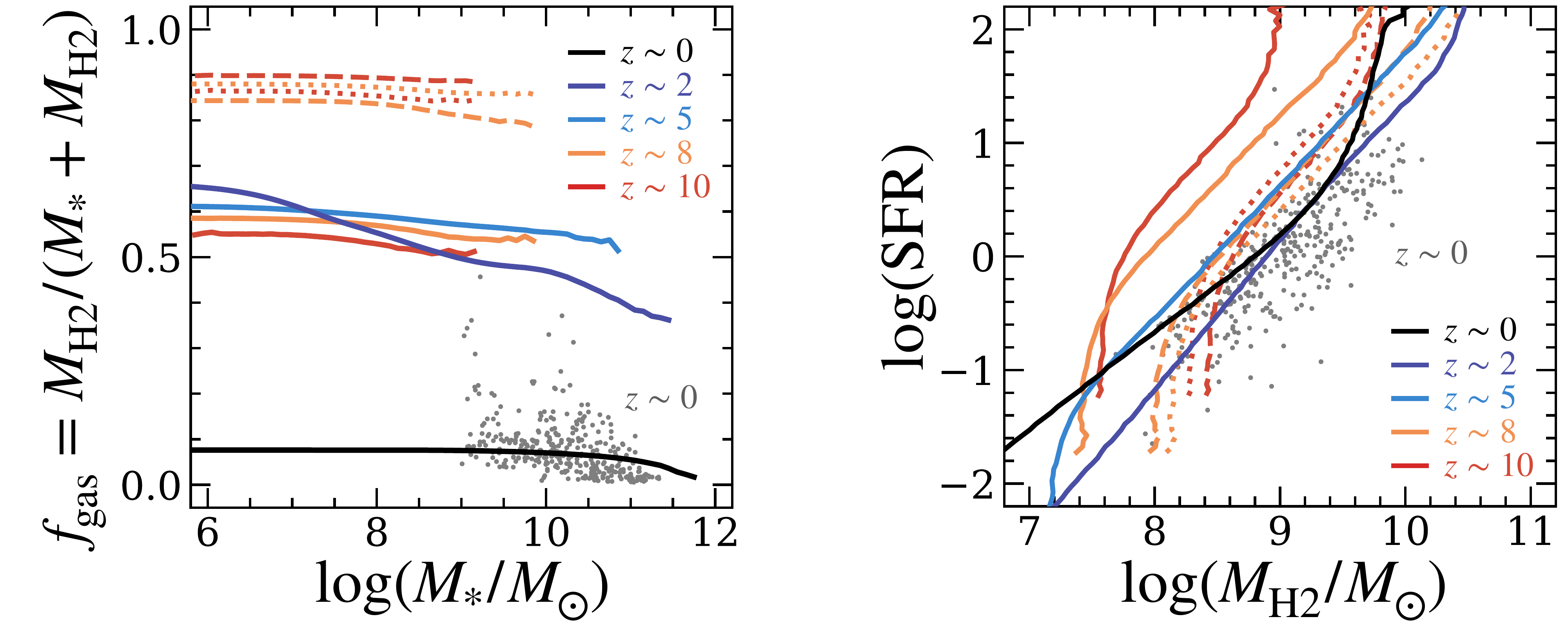}
    \caption{
    Predictions for the evolution of gas properties in the \textsc{ChemicalUniverseMachine}.
    \textbf{Left:} The gas fraction, defined as $f_{\mathrm{gas}} = M_{\mathrm{H2}} / (M_{\mathrm{H2}} + M_*)$, as a function of stellar mass. Gray dots represent local galaxies from the xCOLD GASS survey \citep{Saintonge17}, which are well reproduced by the model at $z \sim 0$ (purple curve). While the standard extrapolation (solid lines) predicts $f_{\mathrm{gas}} \sim 0.5$--$0.7$ at $z > 5$, the scenarios that reproduce the FMR break (Scenarios $b$ and $b^\prime$; dashed and dotted lines) imply extremely high gas fractions of $f_{\mathrm{gas}} \approx 0.8$--$0.9$.
    \textbf{Right:} The relationship between star formation rate and molecular gas mass ($M_{\mathrm{H2}}$). Colors and symbols are the same as in the left panel. The standard extrapolation implies an increase in star formation efficiency (SFR/$M_{\mathrm{H2}}$) with redshift. In contrast, Scenarios $b$ and $b^\prime$ remain consistent with the relation observed at $z \sim 0$–$5$, suggesting little to no evolution in the star formation law.
    }
    \label{fig:fgas}
\end{figure*}
To further explore this scenario, we employ the \textsc{ChemicalUniverseMachine} model \citep{Nishigaki25}. This model builds upon the \textsc{UniverseMachine} \citep{Behroozi19} and \textsc{NeutralUniverseMachine} \citep{Guo23} framework, modeling the evolution of galaxy metallicity by parameterizing how metals are distributed between molecular ($\mathrm{H_2}$) and atomic (HI) gas phases in the ISM.
A key parameter in this framework is the $\mathrm{H_2}$ distribution fraction, $f_\mathrm{H2}$. This is defined as the fraction of metal mass produced by star formation (and ejected by supernovae) that is retained within the molecular gas ($\mathrm{H_2}$) phase. This fraction accounts for both the metals that remain resident in the ISM and those that are temporarily expelled but subsequently recycled via galactic fountains. The remaining metals, $1-f_\mathrm{H2}$, are distributed to the atomic gas (HI), the circumgalactic medium (CGM), or lost to the IGM.
This model has been demonstrated to self-consistently reproduce the observed metallicity relations in the ISM, damped Lyman-alpha systems, and the CGM at $z < 5$. Since the model constrains average relations between stellar mass, SFR, and metallicity for a given halo mass and redshift, rather than tracing individual halo histories, we further assume for the purpose of FMR comparison that $f_\mathrm{H2}$ is a function solely of halo mass and redshift, independent of the specific star formation rate.

However, extending this analysis to the high-redshift regime ($z > 5$) requires extrapolation, as the gas mass fitting functions in the underlying \textsc{NeutralUniverseMachine} (NUM) are calibrated only at $z\lesssim5$. In the framework of this model, the evolution of the FMR is governed by three primary factors: the star formation rate, the gas inflow rate (traced by the molecular gas mass, $M_{\mathrm{H2}}$), and the outflow efficiency (parameterized by $f_{\mathrm{H2}}$). Since the SFR evolution is already constrained by the \textsc{UniverseMachine}, we explore potential evolutionary pathways for the remaining two components, gas mass and metal retention.

We define a baseline using the NUM-based extrapolation, which extends the scaling relations calibrated in the \textsc{NeutralUniverseMachine} at $z < 5$ directly out to $z > 5$.
In this standard model, the star formation efficiency (SFE; defined as $\mathrm{SFR}/M_{\mathrm{H2}}$) increases with redshift, while the gas mass ($M_{\mathrm{H2}}$) remains roughly constant at fixed halo mass. To investigate the origin of the observed FMR deviation, we compare this baseline with three alternative scenarios for $z > 5$.

The first is the Outflow Enhanced scenario (Scenario $a$), where we assume that the metal distribution fraction ($f_{\mathrm{H2}}$) decreases from $z \sim 6$ to $z \sim 8$, representing stronger metal loss (Figure \ref{fig:scenarios}, Left). To reproduce the observed FMR deviation at $z \sim 8$, $f_{\mathrm{H2}}$ is required to be $\sim$0.6 times the value predicted by the standard extrapolation, which corresponds to roughly a factor of $\sim$1.7 increase in outflow efficiency relative to the baseline.

The second is the Inflow Enhanced scenario (Scenario $b$), where we modify the evolution of $M_{\mathrm{H2}}$ such that it increases beyond the standard extrapolation between $z \sim 6$ and $z \sim 8$, also adjusted to match the observations (Figure \ref{fig:scenarios}, Middle). In this scenario, the gas mass at $z \sim 8$ is $\sim$5 times higher than the value predicted by the standard extrapolation, implying an increase by a factor of five in inflow efficiency. 

The third is the Constant SFE scenario (Scenario $b^\prime$), which serves as a physically motivated variation for increasing the gas mass. This model assumes that the star formation efficiency remains constant at $z > 5$ (Figure \ref{fig:scenarios}, Right). Because the SFR increases with redshift at fixed halo mass, this assumption inherently implies that $M_{\mathrm{H2}}$ also increases, resulting in a gas-rich state. Specifically, this leads to a gas mass approximately 3 times higher than the standard extrapolation, corresponding to an increase by a factor of three in inflow efficiency.

We compare the predictions of these scenarios with our observational results in Figure \ref{fig:fmr_um}. The standard NUM-based extrapolation (black curve) predicts no significant deviation from the FMR even at $z > 8$, failing to reproduce the observed decline in metallicity. In contrast, Scenarios $a$, $b$, and $b^\prime$ all predict lower metallicities at higher redshifts, showing good agreement with the observed trend. Here, we examine the behavior of the model parameters in these successful scenarios to understand their physical implications.

Scenario $a$ assumes gas masses and star formation efficiencies unchanged from the standard model but reproduces the observations by assuming a decrease in the metal distribution fraction ($f_{\mathrm{H2}}$). Physically, this implies that metal loss is the dominant mechanism. A lower $f_{\mathrm{H2}}$ signifies that a smaller fraction of the metals produced by supernovae is retained within the star-forming regions. This suggests that metals are being more efficiently ejected into the CGM or IGM, pointing to enhanced outflow efficiency in the early universe compared to lower redshifts.

In contrast, Scenarios $b$ and $b^\prime$ are characterized by a significant increase in gas mass ($M_{\mathrm{H2}}$) at fixed halo mass compared to the standard extrapolation, while the metal distribution fraction remains unchanged.
To quantify the gas properties implied by these scenarios, we present the evolution of the gas fraction and the star formation law in Figure \ref{fig:fgas}.
The left panel shows the gas fraction, $f_{\mathrm{gas}} = M_{\mathrm{H2}}/(M_{\mathrm{H2}}+M_*)$. At high redshifts ($z > 5$), the standard extrapolation predicts gas fractions of $f_{\mathrm{gas}} \approx 0.5$--$0.7$. However, Scenarios $b$ and $b^\prime$, which successfully reproduce the FMR break, require significantly higher values of $f_{\mathrm{gas}} \approx 0.8$--$0.9$. This indicates that galaxies in these epochs are likely extremely gas-dominated systems.
The right panel shows the relation between SFR and $M_{\mathrm{H2}}$. The standard extrapolation (solid lines) implies a shift toward higher SFR at fixed gas mass, indicating an increase in SFE. Conversely, Scenarios $b$ and $b^\prime$ align closely with the relation at $z \sim 0$ even out to $z \sim 10$.
This constancy implies that the local physical processes governing the conversion of molecular clouds into stars have not fundamentally changed from lower redshifts. Consequently, the primary driver of the observed state in these scenarios is not a change in the mode of star formation, but rather the availability of fuel. The presence of a larger reservoir of gas relative to the stellar mass dilutes the concentration of metals produced by star formation. Such a gas-rich state corresponds to a scenario where the inflow of pristine gas is significantly enhanced in the early universe, leading to an accumulation of gas that drives high star formation rates while simultaneously lowering the metallicity.
This picture is consistent with ALMA \CII\ observations of $z \sim 5$ galaxies, which indicate that gas fractions increase with redshift, while the star formation law remains similar to that in the local universe \citep{Dessauges-Zavadsky20,Schaerer20}.

These comparisons demonstrate that to explain the breakdown of the FMR at high redshifts, the physical conditions of galaxies should deviate from the standard extrapolation in two ways. 
One possibility is enhanced outflows, which lead to a more efficient removal of metals from the galaxy, as represented by Scenario ($a$). 
Alternatively, the deviation can be explained by enhanced inflows, resulting in a stronger dilution of the ISM driven by larger gas reservoirs (Scenarios $b$ and $b^\prime$). 
While it is difficult to definitively distinguish between these mechanisms with current data, our analysis with the \textsc{ChemicalUniverseMachine} suggests that the early universe is in a non-equilibrium state driven by one, or a combination, of these evolving physical processes.
\subsection{Suppression of Pristine Gas Supply by Reionization} 
An alternative explanation involves the impact of cosmic reionization. Before reionization is complete, low-mass ($M_{\rm h} \lesssim 10^8M_\odot$) dwarf galaxies, which are predicted to host low-metallicity gas if following a mass–metallicity relation, merge into larger systems and supply pristine gas via minor mergers. However, as reionization proceeds and the UV background is established, halos below $10^{8} M_\odot$ can no longer retain their gas \citep[e.g.,][]{Foreman13,Hasegawa13}. Consequently, subsequent minor mergers would predominantly involve gas-poor halos, 
reducing the efficiency of metallicity dilution through accretion of pristine gas. In the context of the \textsc{ChemicalUniverseMachine} model, however, it is difficult to directly test this scenario because the resolution of the underlying dark matter simulations limits the analysis to halos with $M_{\rm h} > 10^{10}M_\odot$.
\begin{table*}[htb]
    \centering
    \caption{Physical properties of individual galaxies at $z>7.8$ used in this study.}
    \label{tab:indiv_9}
    \begin{tabular}{lccccc}
        \hline \hline
        Object name & Redshift & $\log(M_*/M_\odot)$ & SFR & $12 + \log(\mathrm{O/H})$ & Ref. \\
         & & & [$M_\odot\,\mathrm{yr}^{-1}$] & & \\
        \hline
        \addlinespace
        CEERS 01027 & 7.82 & 8.11 & 44 & 8.00 & (1) \\
        DREAMS 31001 & 7.85 & 7.73 & 5.5 & 7.34 & (1) \\
        DREAMS 40005 & 7.85 & 6.59 & 0.4 & 8.00 & (1) \\
        JADES 21842 & 7.98 & 8.34 & 6.7 & 7.73 & (1) \\
        JADES 1899 & 8.28 & 8.90 & 12 & 8.00 & (1) \\
        DREAMS 30002 & 8.34 & 7.73 & 4.6 & 8.00 & (1) \\
        DREAMS 30003 & 8.34 & 8.67 & 5.4 & 7.98 & (1) \\
        JADES 20213084 & 8.49 & 7.70 & 35 & 7.80 & (1) \\
        ERO 04590 & 8.50 & 7.60 & 12 & 6.96$^{\hyperlink{note:a}{a}}$ & (1) \\
        CEERS 01019 & 8.68 & 9.97 & 126 & 7.87 & (1) \\
        JADES 265801 & 9.43 & 8.51 & 12 & 7.43$^{\hyperlink{note:a}{a}}$ & (1) \\
        \addlinespace
        \hline
        \addlinespace
        RXJ2129-z95 & 9.51 & 7.63 & 1.7 & 7.48$^{\hyperlink{note:b}{b}}$ & (2) \\
        MACS0647-JD & 10.17 & 8.10 & 5.0 & 7.79$^{\hyperlink{note:a}{a}}$ & (3) \\
        GNz-11 & 10.60 & 8.73 & 24 & 7.91$^{\hyperlink{note:a}{a}}$ & (4) \\
        GS-z11-1 & 11.28 & 7.84 & 2.1 & 7.50$^{\hyperlink{note:c}{c}}$ & (5) \\
        GHZ2 & 12.33 & 9.03 & 7.0 & 7.40$^{\hyperlink{note:d}{d}}$ & (6) \\
        \addlinespace
        \hline
        \addlinespace
    \end{tabular}
    \parbox{0.9\linewidth}{
    \footnotesize
    \textbf{Notes.} The listed properties are for two distinct samples of high-redshift galaxies.
    The galaxies at $z=7.8-9.5$ are part of our $z\sim8$ stacked sample. The stellar masses, SFRs, and oxygen abundances are derived in this work.
    The galaxies at $z>9.5$ are not included in our stacking analysis as their \OIIIlam{5007} line is redshifted out of the NIRSpec wavelength coverage. The properties are compiled from the literature.
    The oxygen abundances ($12+\log(\mathrm{O/H})$) for individual galaxies in our stack sample are derived using the R23 calibration from \citet{Nakajima22b}, except for annotated objects for which we use the direct temperature ($T_e$) method.
    For the $z>9.5$ literature sample, the metallicity derivation methods are adopted from their respective original publications:
    \hypertarget{note:a}{($a$)} direct temperature ($T_e$) method,
    \hypertarget{note:b}{($b$)} R23 and O32 calibrations,
    \hypertarget{note:c}{($c$)} Ne3O2 calibration, and
    \hypertarget{note:d}{($d$)} R3 calibration.
    \\
    \textbf{References.} (1) This work; (2) \cite{Williams23}; (3) \cite{Hsiao24}; (4) \cite{Alvarez-Marquez25}; (5) \cite{Scholtz25}; (6) \cite{Zavala24}.
    }
\end{table*}
\begin{table*}[htb]
    \centering
    \caption{Galaxy properties derived from our stacked spectra using the entire sample in each redshift bin, and a mean of individual $z > 9.5$ galaxies from the literature.}
    \label{tab:props}
    \begin{tabular}{lccccccc}
        \hline \hline
        Sample & Redshift & $N$ & $\log(M_*/M_\odot)$ & SFR & $12 + \log(\mathrm{O/H})$ & $\log U$ & $T_\mathrm{e}$ \\
         & & & & [$M_\odot\,\mathrm{yr}^{-1}$] & & & [$10^4$\,K] \\
        \hline
        $z\sim3$ stack & 2–4 (3.09) & 122 & 9.16 & 29 & $8.07^{+0.08}_{-0.09}$ & $-2.61^{+0.03}_{-0.04}$ & $1.29^{+0.09}_{-0.10}$ \\
        $z\sim4$ stack & 4–5.5 (4.41) & 60 & 8.93 & 20 & $7.96^{+0.10}_{-0.12}$ & $-2.38^{+0.04}_{-0.04}$ & $1.40^{+0.13}_{-0.12}$\\
        $z\sim6$ stack & 5.5–7 (6.06) & 49 & 8.67 & 59  & $7.78^{+0.10}_{-0.12}$ & $-2.14^{+0.09}_{-0.09}$ & $1.58^{+0.15}_{-0.15}$ \\
        $z\sim8$ stack & 7–9.5 (7.85) & 18 & 8.15 & 18 & $7.50^{+0.10}_{-0.09}$ & $-2.10^{+0.13}_{-0.13}$ & $1.96^{+0.16}_{-0.17}$ \\
        $z > 9.5$ mean & 10.60 & 5 & 7.95 & 1.62 & 7.78 & – & - \\
        \hline
        \addlinespace
    \end{tabular}
    \parbox{0.9\linewidth}{
    \footnotesize
    \textbf{Notes.} The Redshift column for our stacked samples indicates the full redshift range and the median redshift in parentheses. The quoted uncertainties represent 16th and 84th percentiles.
    }
\end{table*}
\section{Summary}\label{sec:summary}
We investigate the nebular properties and chemical evolution of star-forming galaxies at redshifts $z = 2-10$ using JWST/NIRSpec medium-resolution spectra from a compilation of public surveys, including DREAMS, JADES, CEERS, and ERO.
By stacking spectra in redshift bins at fixed stellar mass and star formation rate, we derive direct metallicities and investigate the evolution of line ratios, ionization parameters, and the mass–metallicity and fundamental relation. 
Our main results are summarized as follows:
\begin{itemize}
    \item The average \OII/\Hb\ ratio decreases by $\sim$ 0.7 dex from $z \sim 3$ to 8 at fixed $M_*$ and SFR, while \OIII/\Hb\ ratio remains nearly constant, indicating a shift in ionization conditions.
    \item We detect the auroral \OIIIlam{4363} line in stacked spectra and derive direct $T_e$-based gas-phase metallicities, finding that high-redshift galaxies lie at the low-metallicity, high-ionization end of the local metallicity–ionization parameter anti-correlation.    
    \item Photoionization modeling demonstrates that the observed redshift evolution in line ratios is driven by decreasing metallicity and increasing ionization parameter,
    indicating that high-redshift galaxies follow the anti-correlation between ionization parameter and metallicity as observed in the local universe.
    \item We find that the mass–metallicity relation declines from $z \sim 3$ to 10 at fixed stellar mass.
    The fundamental metallicity relation (FMR) appears to evolve at higher redshifts, showing a significant deviation from the local FMR at $z\gtrsim8$.
    \item The \textsc{ChemicalUniverseMachine} predicts no deviation from the FMR even beyond $z\gtrsim8$ when utilizing a standard extrapolation of low-redshift relations. However, scenarios incorporating either enhanced gas inflows (leading to higher gas fractions and dilution) or increased outflow efficiency (leading to reduced metal retention) successfully reproduce the observed FMR breakdown. This suggests that the baryon cycle in the early universe ($z \gtrsim 8$) is likely in a non-equilibrium state, distinct from the local universe, driven by these evolving physical processes.
\end{itemize}
\section*{Acknowledgements}
We thank 
Yusei Koyama,
Yuich Matsuda,
Koichiro Nakanishi,
Ryota Ikeda,
Tomokazu Kiyota,
Yuma Sugahara,
Kazunori Kohri,
Tomohiro Yoshida,
and Hideko Nomura,
for useful comments and discussions that improved our manuscript.
This work is based on observations made with the NASA/ESA/CSA James Webb Space Telescope. The data were obtained from the Mikulski Archive for Space Telescopes at the Space Telescope Science Institute, which is operated by the Association of Universities for Research in Astronomy, Inc., under NASA contract NAS 5-03127 for JWST.
These observations are associated with JWST programs 1180 and 1181 (JADES), 1345 (CEERS), 2736 (ERO), and 4750 (DREAMS). 
We are grateful to all teams for their extensive efforts in developing and conducting these observation programs. We thank the JADES team for the public release of their high-quality data products, including reduced spectra and catalogs.
This paper is supported by World Premier International Research Center Initiative (WPI Initiative), MEXT, Japan, as well as the joint research program of the Institute of Cosmic Ray Research (ICRR), the University of Tokyo. We acknowledge support from JSPS KAKENHI Grant: JP20H00180, JP21H04467, JP24K07102, and 25H00674.
M.N. is supported by JST, the establishment of university fellowships towards the creation of science technology innovation, Grant Number JPMJFS2136. 
H.Y. acknowledges support from JSPS KAKENHI Grant Number 21H04489 and JST FOREST Program Grant Number JP-MJFR202Z.
HY acknowledges support by KAKENHI (25KJ0832) through Japan Society for the Promotion of Science (JSPS).
\bibliographystyle{apj}
\bibliography{MAIN.bib}
%
\end{document}